\documentclass[twocolumn,showpacs,preprintnumbers,amsmath,amssymb]{revtex4}
\begin{document}
\newcommand{\beq}{\begin{equation}}
\newcommand{\eeq}{\end{equation}}
\newcommand{\beqn}{\begin{eqnarray}}
\newcommand{\eeqn}{\end{eqnarray}}
\newcommand{\bmath}{\begin{subequations}}
\newcommand{\emath}{\end{subequations}}
\title{ Electronic dynamic Hubbard model: exact diagonalization study}
\author{J. E. Hirsch }
\address{Department of Physics, University of California, San Diego\\
La Jolla, CA 92093-0319}

\date{\today} 

\begin{abstract} 
A model to describe electronic correlations in energy bands is considered. The model
is a generalization of the conventional Hubbard model that allows for the fact that the 
wavefunction for two electrons occupying the same Wannier orbital
 is different from the product of single
electron wavefunctions. We diagonalize the Hamiltonian exactly on a four-site cluster
and study its properties as a function of band filling. The quasiparticle weight is
found to decrease and the quasiparticle effective mass to increase as 
the electronic band filling increases, and spectral weight in one- and two-particle
spectral functions is transfered from low to high frequencies as the band
filling increases. Quasiparticles at the Fermi energy are found to be more 'dressed'
when the Fermi level is in the upper half of the band (hole carriers) than when it is in the lower half
of the band (electron carriers). The effective interaction between carriers 
is found to be
strongly dependent on band filling becoming less repulsive as the band filling
increases, and attractive near the top of the band in certain parameter ranges.
The effective interaction is most attractive when the single hole carriers are most heavily
dressed, and in  the parameter regime where the effective interaction is attractive, 
hole carriers are
found to 'undress', hence become more like electrons, when they pair.
It is proposed that these are generic properties of electronic energy bands
in solids that reflect a fundamental electron-hole asymmetry of condensed matter.
The relation of these results to the understanding of superconductivity
in solids is discussed.
\end{abstract}
\maketitle 
\vskip2pc]
 
\section{Introduction}
The fact that the Coulomb repulsion between two electrons in a doubly occupied
atomic orbital is smaller than the value predicted from the expectation value of 
the Coulomb interaction with single electron wavefunctions establishes that
atomic orbitals are not infinitely rigid\cite{slater}: electrons will develop intra-atomic
correlations to reduce their Coulomb repulsion. This well-known fact is not
incorporated in the conventional single band Hubbard model\cite{hubbard}, which assumes that
two electrons of opposite spin will doubly-occupy the same single electron atomic orbital.
Dynamic Hubbard models\cite{atom} attempt to remedy this deficiency by either introducing an auxiliary
boson degree of freedom to mimic the orbital relaxation that occurs on double
atomic occupancy, or by allowing more than one atomic orbital. Here we consider
a purely electronic dynamic Hubbard model with  two orbitals per site and no
auxiliary boson degree of freedom.

The model considered here has some superficial resemblance to various multi-orbital
Hubbard models that have been considered in the past such as the degenerate
Hubbard model\cite{deghub}, the Falicov-Kimball model\cite{falkim} and the Anderson lattice 
model\cite{anderson}.
However conceptually it is rather different. The goal is not to model the physics
of electrons in degenerate atomic orbitals, nor of two different partially occupied
bands in a solid, nor of local magnetic moments interacting with conduction electrons,
nor of mixed valence.
Rather, we are interested in modeling the physics of a single band, and based on 
general arguments\cite{atom} argue that it is necessary, to understand single band physics,
to include at least one other higher-lying orbital besides the one that is being filled as the
electronic band is being filled. This second orbital becomes increasingly important
as the filling of the band increases beyond one half.

When an electronic energy band is less than half-full, electrons can avoid each other
by developing elaborate interatomic correlations. For this situation the ordinary
single band Hubbard model may be adequate. However, when the band is more than
half-filled, some Wannier orbitals are $necessarily$ doubly occupied. The resulting large
cost in Coulomb energy cannot be avoided by $any$ interatomic correlations;
instead, it can be reduced by the electrons developing intra-atomic correlations.
The conventional single band Hubbard model cannot describe this, hence it is
inadequate to describe electronic energy bands that are more than
half-full: 
 it is necessary to introduce another degree of freedom. 
Here this is achieved by having a second atomic orbital in
the model, that becomes increasingly occupied as the band filling increases beyond half
filling. As we will see in this paper, the resulting electron-hole asymmetry in the band causes the
quasiparticle weight of carriers in the upper half of the band (holes) to be 
smaller than that in the lower half of the band (electrons), and their effective mass
to be larger. We believe that this physics is generally a part of the physics
of electronic energy bands in solids. Its quantitative importance will depend on
the particular solid under consideration.

In section II we introduce the  electronic dynamic Hubbard model to 
be studied in this paper. Section III gives analytic results in limiting cases,
and section IV presents numerical results for effective interaction, quasiparticle
weight and  effective mass  as a function of band filling and
model parameters. Section V present results for optical absorption, and in
Sect. VI we discuss the relation between the parameters in the Hamiltonian and
atomic parameters in  materials. We conclude in Sect. VII with a discussion of the 
applicability of these results to the description of real electrons in real solids.

\section{Model Hamiltonian} 
The static (conventional) single band Hubbard model is defined by the Hamiltonian
\beq
H=-\sum_{i,j,\sigma} t_{ij}c_{i\sigma}^\dagger c_{j\sigma}+U\sum_i
n_{i\uparrow}n_{i\downarrow}
\eeq
which properly emphasizes the fact that the intra-atomic electron-electron
repulsion is the dominant source of electronic correlation in solids.
Here, $c_{i\sigma}^\dagger$ creates an electron in a Wannier orbital centered at lattice
site $i$, and the hopping amplitude $t_{ij}$ is the Fourier transform of the
single electron band energy $\epsilon_k$:
\beq
t_{ij}=\frac{1}{N}\sum_k e^{ik(R_i-R_j)}\epsilon_k .
\eeq

The fundamental problem with the Hubbard Hamiltonian Eq. (1) is that it implicitly assumes that
the state of two electrons in a Wannier orbital is a single Slater determinant  $c_{i\uparrow}^\dagger  c_{i\downarrow}^\dagger |0>$,
with $|0>$ the state of the empty orbital. This is incorrect, precisely because of the existence of the
electron-electron interaction described by the Hubbard $U$. We extend the Hamiltonian Eq. (1) to allow for the fact that when two electrons
occupy the same Wannier orbital, intra-orbital electronic correlations will develop
due to the large local electron-electron repulsion. This effect can be
described by having at least two Wannier orbitals per site, which the electrons
will partially occupy to reduce their Coulomb repulsion. Let $c_{i\sigma}^\dagger$ and
$c^{'\dagger}_{i\sigma}$
be the operators creating electrons into these Wannier 
orbitals. We consider the local Hamiltonian\cite{old}
\bmath
\beqn
H_i&=&Un_{i\uparrow}n_{i\downarrow}+U'n'_{i\uparrow}n'_{i\downarrow}
+\epsilon n'_i +V n_i n'_i \\ \nonumber
&&-t'\sum_\sigma (c_{i\sigma}^\dagger c'_{i\sigma} + h.c.)
\eeqn 
with $n_{i\sigma}=c_{i\sigma}^\dagger c_{i\sigma}$, $n'_{i\sigma}=
c^{'\dagger}_{i\sigma} c'_{i\sigma}$, $n_i=n_{i\uparrow}+n_{i\downarrow}$,
$n'_i=n'_{i\uparrow}+n'_{i\downarrow}$. In the extreme tight binding limit the
Wannier orbitals become atomic orbitals, and we will cast the discussion in
that language even though it should be more generally applicable.

The primed orbital is higher in energy by $\epsilon$ than the unprimed orbital,
hence more extended in space, so that the Coulomb repulsion $U'$ should be
smaller than $U$. The orbital energy $\epsilon$ will generally
satisfy the relation $t<\epsilon<U$, so that when the site is doubly occupied
electrons will partially occupy this orbital to reduce their Coulomb repulsion
and there is no justification for neglecting this second orbital as is done
in the Hamiltonian Eq. (1). $V$ describes the Coulomb repulsion between one
electron in each orbital, and $t'$ the intra-atomic hybridization between these
orbitals. The lattice Hamiltonian is then
\beqn
H=&-&\sum_{ij} [t_{ij}c_{i\sigma}^\dagger c_{j\sigma}+
t'_{ij}(c_{i\sigma}^\dagger c'_{j\sigma} +h.c.) +t''_{ij}c^{'\dagger}_{i\sigma} c'_{j\sigma}]
\\ \nonumber
&+&\sum_i H_i
\eeqn
\emath
with $H_i$ given by Eq. (3a).  
We expect the various hopping matrix elements 
$t_{i}$, $t'_{ij}$, $t''_{ij}$ to be similar in magnitude and hence will assume for
the remainder of this paper
\beq
t_{ij}=t'_{ij}=t''_{ij}
\eeq
and furthermore that the hopping connects only nearest neighbor sites,
hence $t_{ij}\equiv t$ for $i,j$ nearest neighbors, zero otherwise.

It should be pointed out that we do not expect the qualitative
physics of the model to depend on the particular choice for the hoppings Eq. (4). For example,
in Ref. \cite{old} the same site Hamiltonian Eq. (3a) was considered with hoppings
$t'_{ij}=t''_{ij}=0$ instead of Eq. (4), and the same qualitative physics was obtained as in the
model considered here. Certainly, quantitative differences will occur depending on these
choices that should be studied in the future. In this connection it is also relevant to note that
in a recent study of the periodic Anderson model, different choices of the hybridization
(whether same site or neighboring sites) were found to not affect the underlying physics\cite{impurity}.

\section{Analytic results}
\subsection {Non-interacting band structure}
The single electron part of the Hamiltonian Eq. (3) is, after Fourier transforming
\bmath
\beq
H=\sum_k 
\left(
 \begin{array}{cc}
c^\dagger_{k\sigma} & c^{'\dagger}_{k\sigma} \\
\end{array}
\right)
 H_k
\left(
 \begin{array}{c}
c_{k\sigma} \\ c^{'}_{k\sigma} \\
\end{array}
\right)
\eeq
\beq
H_k=
\left(
 \begin{array}{cc}
\epsilon_k&\epsilon_k-t'\\\epsilon_k-t'&\epsilon_k+\epsilon\\
\end{array}
\right)
\eeq
\emath
with
\beq
\epsilon_k=-2tcosk
\eeq
for a one-dimensional geometry. Two bands result, with energy versus $k$ relation
\beq
E_k^{1,2}=\epsilon_k+\frac{\epsilon}{2} \pm \sqrt{(\frac{\epsilon}{2})^2+
(\epsilon_k-t')^2}
\eeq
The two bands are separated by an indirect gap
\beqn
\Delta&=&E_{k=0}^2-E_{k=\pi}^1=\sqrt{(\frac{\epsilon}{2})^2+(2t-t')^2}+
\\ \nonumber
& &\sqrt{(\frac{\epsilon}{2})^2+(2t+t')^2} -4t
\eeqn
which is always positive for $\epsilon>0$. The bandwidth of the two bands
is approximately given by $D_1=4t+2t'$, $D_2=4t-2t'$. Figure 1 shows the
band structure for various parameters. Note that the effective mass in the
lower band is larger near the top of the band than near the bottom;
this effect becomes much more pronounced in the presence of electron-electron
interactions.

The zero temperature single electron spectral function for the system with
$N$ electrons is 
\beqn
A_\alpha(&k&,\omega)=\sum_n |<n_{N+1}|c_{\alpha k \sigma}^\dagger|0_N>|^2 \times
\\ \nonumber
& &\delta(\omega-(E_n^{N+1}-E_0^{N+1}+\mu_N))+\\ \nonumber
& &
|<n_{N-1}|c_{\alpha k \sigma}|0_N>|^2
\delta(\omega+(E_n^{N-1}-E_0^{N-1}-\mu_{N-1}))
\eeqn
Here, $|n_N>|$ is the n-th excited state of the system with
$N$ electrons ($n=0$ is the ground state) and $E_n^N$ is the
eigenvalue, and $\mu_N=E_0^{N+1}-E_0^N$, $\mu_{N-1}=E_0^{N}-E_0^{N-1}$.
For a metal $\mu_N=\mu_{N-1}\equiv \mu$ and we redefine the frequency
$\omega\rightarrow \omega+\mu$ so that  
\beqn
A_\alpha(k,\omega)&=&\sum_n |<n_{N+1}|c_{\alpha k \sigma}^\dagger|0_N>|^2 \times \\ \nonumber
& &\delta(\omega-(E_n^{N+1}-E_0^{N+1}))+\\ \nonumber
&|&<n_{N-1}|c_{\alpha k \sigma}|0_N>|^2
\delta(\omega+(E_n^{N-1}-E_0^{N-1}))
\eeqn
The index $\alpha$ labels the fermion operator of the unprimed or primed orbital,
or an appropriate linear combination theerof. The creation operator for an electron in
band 1 is given by
\beq
c_{1k\sigma}=u_kc_{k\sigma}+v_kc'_{k\sigma}
\eeq
with $u_k$, $v_k$ obtained from diagonalization of $H_k$, Eq. (5b).
Using this operator in Eq. (10) we have simply
\beq
A_1(k,\omega)=\delta(\omega-(E_k^1-\mu))
\eeq
and the quasiparticle weight for a single electron in this band is the
coefficient of the $\delta$-function in Eq. (12)
\beq
z_k=1
\eeq

\subsection{Quasiparticle weight for interacting system}

In an interacting many-body system the spectral function has the
form
\beq
A_\alpha(k,\omega)=z_k\delta(w-(\epsilon_k-\mu))+A'_\alpha(k,\omega)
\eeq
where $0\leq z_k \leq 1$ is the quasiparticle weight, and
$A'_\alpha$ is the incoherent part of the spectral function.
We define the quasiparticle weight in our model at the Fermi
energy by
\beq
z(N)=|<O_{N-1}|c_{\alpha k_F\sigma}|O_N>|^2
\eeq
with the quasiparticle operator defined by
\beq
c_{\alpha k_F\sigma}=u_{k_F}c_{k_F\sigma}+v_{k_F}c'_{k_F\sigma}
\eeq
with $u_k^2+v_k^2=1$. We choose the particular linear combination that
maximizes the quasiparticle weight Eq. (15). The following simple relations
are easily proven:
\beq
z(N)=\sum_i[|<O_{N-1}|c_{i\sigma}|O_N>|^2+
|<O_{N-1}|c'_{i\sigma}|O_N>|^2]
\eeq
\bmath
\beq
u_{k_F}^2=\frac{\sum_i|<O_{N-1}|c_{i\sigma}|O_N>|^2}{z(N)}
\eeq
\beq
v_{k_F}^2=\frac{\sum_i|<O_{N-1}|c'_{i\sigma}|O_N>|^2}{z(N)}
\eeq
\emath
Eq. (17) gives the total quasiparticle weight at the Fermi energy, and
$1-z$ gives the amount of spectral weight in the incoherent
part of the spectral function. The quantities $u_{k_F},v_{k_F}$ indicate
how much of the quasiparticle resides in the lower and upper orbital
respectively. For a single band conventional Hubbard model the
quasiparticle weight at the Fermi energy is given by the first
term in Eq. (17).

\subsection{Optical conductivity and effective mass}
The current operator in our model is given by
\beq
J=it\sum_i[(c_{i+1\sigma}^\dagger +c_{i+1\sigma}^{'\dagger})
(c_{i\sigma}  +c'_{i\sigma})-h.c.]
\eeq
and we will compute the optical conductivity at zero temperature given by
\beq
\sigma_1(\omega)=\pi\sum_m\frac{|<0|J|m>|^2}{E_m-E_0}\delta(\omega-(E_m-E_0))
\eeq
The optical sum rule states that the integral of the optical
conductivity is\cite{maldague}
\beq
\int_0^\infty d\omega \sigma_1(\omega)=\frac{\pi}{2}<0|-T|0>
\eeq
with the kinetic energy given by

\beq
T=-t\sum_{i\sigma\delta}(c_{i\sigma}^\dagger+c_{i\sigma}^{'\dagger})
(c_{i+\delta\sigma}+c'_{i+\delta\sigma})
\eeq
In our two-orbital model, optical transitions include
both 'intraband' transitions as well as interband transitions to the
second band. For parameters where a clear separation of energy scales
occurs one can define an effective Hamiltonian describing the low
energy part of the Hilbert space, and write a 'partial'
conductivity sum rule
\beq
\int_0^{\omega_m} d\omega \sigma_1(\omega)=A_l=\frac{\pi}{2}<0|-T_{eff}|0>
\eeq
where $T_{eff}$ is the kinetic energy in the lower band, and the high 
frequency cutoff $\omega_m$ excludes transitions to the second band.

\subsection{Strong coupling limit}
We consider the Hamiltonian Eq. (3) in the parameter regime
\bmath
\beq
U'+2\epsilon<V+\epsilon<U
\eeq
\beq
U,U',V>>\epsilon>>t'
\eeq
\emath
These conditions ensure that a single electron at a site resides 
primarily in the lower orbital, while in the doubly occupied site two
electrons of opposite spin reside primarily in the higher orbital. Some results in 
this limit were discussed in ref. \cite{old}. The site eigenstates
to lowest order in $t'$ are
\bmath
\beq
|\tilde{\uparrow}>=|\uparrow>|0>+\delta|0>|\uparrow>
\eeq
\beq
|\tilde{\uparrow \downarrow}>=|0>|\uparrow \downarrow>+
\delta'[|\uparrow>|0>+|0> \downarrow>|0>]
\eeq
\emath
where the first (second) ket in the product refers to the lower
(higher) orbital, and
\bmath
\beq
\delta=t' / \epsilon
\eeq
\beq
\delta'=\frac{t'}{V-U'-\epsilon}
\eeq
\emath
The single site quasiparticle weight for a hole is, from Eq. (17)
\bmath
\beq
z_h=|<\tilde{\downarrow}|c_\uparrow|\tilde{\uparrow\downarrow}>|^2+
|<\tilde{\downarrow}|c'_\uparrow|\tilde{\uparrow\downarrow}>|^2
=(\delta+\delta')^2\equiv S^2
\eeq
and the single site quasiparticle weight for an electron is
\beq
z_e=|<0|c_\uparrow|\tilde{\uparrow}>|^2+
|<0|c'_\uparrow|\tilde{\uparrow}>|^2=1
\eeq
\emath
Hence the single site spectral function for a  hole has other terms in
addition to the quasiparticle term since $z_h<1$, while the 
site spectral function for an electron is a single $\delta-$function.
Consequently, in the solid the spectral function for a single
electron in the ground state of the Hamiltonian Eq. (3) is a
$\delta-$function, while the spectral function for a single hole
in the lower band has incoherent contributions, as discussed
in ref. \cite{atom}.

We can define quasiparticle operators for the lower band
$\tilde{c}_{i\sigma}$ as operators connecting the site ground states
\bmath
\beq
\tilde{c}_{i\sigma}^\dagger|0>=|\tilde{\sigma}>
\eeq
\beq
\tilde{c}_{i\sigma}^{'\dagger}|0>=|\tilde{\uparrow\downarrow}>
\eeq
\emath
and the relation between bare electron and quasiparticle
operators is\cite{undressing}
\beq
c_{i\sigma}=[1+(S-1)\tilde{n}_{i,-\sigma}]\tilde{c}_{i\sigma}
\eeq
Replacing the bare fermion operators in the kinetic energy
Eq. (3b) in terms of the quasiparticle operators yields an
effective Hamiltonian for quasiparticles in the lower band, of the form
\beqn
H_{eff}&=&-t\sum_{ij,\sigma}[1+(S-1)(\tilde{n}_{i,-\sigma}+
\tilde{n}_{j,-\sigma})+\\ \nonumber & &(S-1)^2\tilde{n}_{i,-\sigma}
\tilde{n}_{j,-\sigma}](\tilde{c}_{i\sigma}^\dagger\tilde{c}_{j\sigma}
+h.c.)
\eeqn
This Hamiltonian gives rise to a 'correlated hopping' term which leads to 
pairing of holes. The enhancement of the hopping amplitude for a hole
when the sites involved in the hopping process have an additional hole
is 
\beq
\Delta t=tS(1-S)
\eeq
and leads to superconductivity as discussed in detail elsewhere\cite{marsiglio}.

\section{Numerical results}
We diagonalize the Hamiltonian Eq. (3) on an $N=4$-site lattice with number of
electrons $N_e$ ranging from $0$ to $8$, as appropriate to fill the lower band.
The maximum number of states in the Hilbert space is $4900$. To reduce the
computational complexity and because we will be interested in the case
of large Coulomb interactions we discard all states where there
are more than two electrons at any given site from the Hilbert space.
We also discard states that have two electrons of the same spin at a site,
because we will not be interested in spin-polarized states. This simplification should not change
the qualitative physics of interest here, and such states certainly become irrelevant
in the limit of large interorbital repulsion $V$; however it could have a quantitative
effect for small $V$ and should be investigated further. With these
simplifications the maximum size of the Hilbert space is $1024$ states when
there are $6$ electrons in the four-site system.

Figure 2 shows schematically the band under consideration and the number of 
states in the Hilbert space for different positions of the Fermi level.
Note that there are many more states in the Hilbert space for a given
number of holes in the band than for the same number of electrons. This fact alone
indicates that the system will be more incoherent when the band is more
than half filled compared to when it is less than half filled. The figure shows
one representative state in each sector: when there is one electron at the
site, it resides primarily in the lower orbital, when there are two electrons
at the site they are depicted as occupying the higher orbital which will be
the dominant contribution in the parameter range of interest.

\subsection{Choice of boundary conditions}
For the small system under study boundary conditions are of course important.
We have calculated properties of the system using periodic, antiperiodic
and free ends boundary conditions (bc). While periodic bc may yield results closer
to the thermodynamic limit in some cases, we believe free ends bc are 
preferable for several reasons. Most importantly, the single particle eigenstates
for the finite chain are non-degenerate with free ends bc, while degenerate
states occur for both periodic and antiperiodic bc.

Consider the effective interaction between two electrons in the chain with
$N_e$ electrons, defined as
\beq
U_{eff}(N_e)=E_0(N_e)+E_0(N_e-2)-2E_0(N_e-1)
\eeq
with $E_0(N_e)$ the ground state energy with $N_e$ electrons.
Figure 3 shows the effective interaction versus band occupation for various parameter
values. For all the different boundary conditions the effective interaction becomes
attractive near the top of the band for certain parameter values. This is a robust
effect. On the other hand, it can be seen that the effective interaction can be
negative for the half-filled band case for periodic bc, and it is zero near the
bottom of the band for antiperiodic bc. These are spurious effects related to
finite system degeneracies, that are expected to dissappear in the thermodynamic
limit\cite{fye}. Instead, for free ends bc the effective interaction is repulsive
except near the top of the band. This behavior should persist for larger systems.
Quantitatively, we expect the magnitude of the attractive interaction to be 
somewhat smaller for the infinite system than for the small cluster\cite{qmc}.

Another reason to use free ends boundary conditions is that the optical sum rule
is satisfied in that case for the finite system\cite{hanke}, while if periodic
boundary conditions are used the 'Drude weight' needs to be added by hand to the
optical response. Furthermore, the expression for the quasiparticle weight
Eq. (17) is not correct when there are degenerate states at the Fermi energy.
Hence we will use free ends boundary conditions for the remainder of this paper.

\subsection{Quasiparticle bands}
We begin our study with a look at the energy level spectrum of the model. We choose 
sets of parameters that yield a clear separation of the spectrum of the 
lower band states. Figure 4 shows the energy levels as a function of $N_e$, the number
of electrons in the cluster, for a non-interacting case (a) and for an
interacting case (b). The dashed line separates the low-lying band 'intraband' states from 
the other states. For the noninteracting case, the spectrum in the lower band
is nearly symmetric for electrons and holes, as one would expect. Instead,
with electron-electron interactions a large electron-hole asymmetry exists.

Consider first the states for a single electron ($N_e=1$) and for a single hole
($N_e=7$). There are 4 states in each case (= the number of sites in the cluster)
and the distance between the lowest and highest intraband state is the bandwidth.
Clearly, for the interacting case the hole bandwidth is much smaller 
than the electron bandwidth.
The effective mass of the quasiparticle is inversely proportional to the spacing
between intraband states. Clearly, the quasihole is substantially heavier than the
quasielectron in the presence of electron-electron interactions for these parameter
values. 

As the number of electrons or of holes is increased, the number of intraband states
increases. There are 6 intra-band states for the half-filled band ($N_e=4$), and 
12 states for all other occupations, because states with double site occupancy are
pushed to much higher energies due to the large Coulomb repulsions. Note the
asymmetry in the spectra in the interacting case for band filling less and more
than one half: the quasiparticle band is substantially narrower when the carriers
are holes (more than half-filled band) compared to the corresponding case when
the carriers are electrons. In the effective strong coupling Hamiltonian
one obtains a quasiparticle bandwidth that decreases monotonically as the electronic
band filling increases:
\beq
D(n_e)=D[1-\frac{n_e}{2}(1-S)]^2
\eeq
with $ 0\leq n_e=N_e/N \leq 2$ the band filling.

\subsection{Quasiparticle properties versus band filling}

We consider first a case where the second band is well separated in energy,
with $\epsilon=10$. Consider the evolution of the quasiparticle weight at the
Fermi energy, given by Eq. (17), as the magnitude of the Coulomb interactions
increase, shown in Fig. 5a. The quasiparticle weight is 1 for the noninteracting case,
and it remains 1 in the presence of interactions at the bottom and at the top of
the band, since the single electron and the single hole behave as free particles. 
When the band filling increases from empty or decreases from full the
quasiparticle weight decreases in the presence of interactions and is lowest at the
half filled band. Thus the spectral function will have largest incoherent
contribution at and close to half-filling, as one expects in the conventional
single band Hubbard model.

The quasiparticle weights in Fig. 5a appear to be electron-hole symmetric. However,
a close look reveals that, except for the noninteracting case, the quasiparticle
weight for holes is always slightly smaller than the corresponding one for
electrons, i.e.
\beq
z(n_e)>z(2-n_e)
\eeq
for $n_e<1$. This effect is due to the presence of the second band, and will exist
always as long as $\epsilon < \infty$ in our model.

Figure 5b shows the effective interaction defined by Eq. (32) versus band filling
for these cases. The effective interaction becomes more repulsive as the bare
repulsion parameters increase, and is approximately electron-hole symmetric
in this case, as in the case of the conventional Hubbard model.

We can estimate the effective mass or the effective hopping amplitude for the
quasiparticle from the difference in energy between the ground state and the
first excited state. Figure 5c shows the results for these cases. For the
noninteracting case $t_{eff}$ is approximately constant versus band filling,
and as the repulsive interactions increase it decreases as the number of
carriers in the band increases. Here the electron-hole asymmetry due to the
fact that $\epsilon<\infty$ is more apparent, with holes always being
heavier than electrons.

Next we consider the effect of decreasing the interband energy separation
$\epsilon$ in the presence of Coulomb repulsion. Figure 6 shows results
for $\epsilon=10,5,4$ and $2$. As $\epsilon$ decreases, the quasiparticle properties
become increasingly electron-hole asymmetric. The quasiparticle weight, shown
in Fig. 6a, is substantially smaller for holes than for electrons as
$\epsilon$ becomes small. Similarly the effective hopping amplitude
shown in Fig. 6c decreases as the band filling increases, and holes are
much heavier than electrons when $\epsilon$ becomes small. The effective
interaction (Fig. 6b) remains repulsive for these parameters for all
band fillings.

For sufficiently small $\epsilon$ and not too large $U'$ however the effective
interaction at the top of the band will become atractive. Figure 7 shows
results for $\epsilon=2$ and $U'=2$. We also show for comparison the cases
$\epsilon=2$, $U'=5$ and $\epsilon=5$,$U'=2$, where the effective 
interaction is always repulsive: both small $\epsilon$ and small
$U'$ are required to yield an attractive interaction for holes.

It is interesting to examine the quasiparticle weight and effective hopping
amplitude for these cases. For the parameters where $U_{eff}$ is attractive
the quasiparticle weight for a single hole is smallest, and the quasiparticle weight
is larger for two holes (Fig 7a). This is in contrast to the cases of repulsive
$U_{eff}$, where the quasiparticle weight for 2 holes is smaller than for
1 hole. Similarly the effective hopping (Figure 7c) for two holes is larger than 
for one hole when $U_{eff}$ is attractive, and smaller when $U_{eff}$ is repulsive.
In other words, quasiparticles 'undress',  i.e. increase their quasiparticle weight
and decrease their effective mass, when they pair.

It is also interesting to examine the expectation value of the kinetic energy
operator, Eq. (22). This is shown in Figure 8 for the three parameter sets
under consideration. The kinetic energy is lowered both when electrons are
added to the empty band and when holes are added to the full band. 
For the case of attractive $U_{eff}$ where the single hole is highly dressed 
(small $z$) and the effective hopping is smallest, the kinetic energy is highest, as 
one would expect. As Fig. 8 shows, in that case only when a second hole is added
to the full band the kinetic energy decreases $below$ twice the value of the
single hole kinetic energy. This indicates that pairing of holes is driven by
lowering of kinetic energy in this model.

Finally, Figure 9 shows the composition of the quasiparticles, as given by
the quantities $u_{k_F}, v_{k_F}$ in Eq. (18). As $\epsilon$ decreases the
quasiparticles occupy predominantly the higher orbital (large $v_{k_f}$) when
the number of electrons in the band increases. In the case where the pairing
interaction is attractive the quasiparticle weight for a hole in the system with
2 holes is also dominantly in the higher orbital, because of the large 
probability for two holes to be on the same site; in contrast, for small
$\epsilon$ and larger $U'$ when the holes are not paired, $v_{k_f}$ is much
smaller for the system with two holes because the holes occupy different sites.
For the case of large $\epsilon$, the quasiparticle weight is dominantly
in the lower orbital for all fillings. These results indicate that a necessary
but not sufficient condition for pairing in this model is that parameters
are such that there is a large probability for electrons to occupy 
the higher orbital when the band is close to full.

\subsection{Hole pairing}

As seen in the previous section, the model can give rise to pairing for
carriers near the top of the quasiparticle band with repulsive Coulomb
interactions. In this section we examine in which regions of parameter space
is the effective interaction between holes attractive, and shed some light
on its origin.

Figure 10 shows the dependence of $U_{eff}$ for two holes at the top of
the band on various Hamiltonian parameters. As seen in Fig. 10a, $U_{eff}$ is attractive when the
repulsive interaction in the higher orbital, $U'$, is sufficiently small. 
This is because two electrons at a site will occupy dominantly the higher atomic
orbital in the regime where pairing occurs. The energy difference between
the two atomic orbitals, $\epsilon$, plays an important role: both for large and
for small $\epsilon$ the attraction is suppressed, with $\epsilon\sim 1.5$ yielding
the largest range of $U'$ where $U_{eff}$ is attractive. This value of 
$\epsilon$ is found to be optimal for a wide range of the parameters
$U$, $V$ and $t'$, i.e. it is set by the value of the hopping amplitude $t=1$.

In contrast, attraction between holes is favored by a large value of the 
lower orbital repulsion $U$, as seen in Figure 10b (we ignore the small region of
very small $U$ where $U_{eff}$ is attractive which is presumably unphysical).
This is because the attraction requires a large change in the state of the
remaining electron when a second electron is removed from the site. If $U$ 
becomes small, two electrons will occupy the smaller rather than the higher
orbital and this effect is lost. Similarly, we find (not shown) that large
values of the interorbital Coulomb repulsion $V$ are favorable to pairing:
the dependence of $U_{eff}$ on $V$ is similar to the dependence on $U$ shown
in Fig. 10b.

As a function of the interorbital hybridization $t'$, pairing will occur when
$t'$ is not too large, as seen in Fig. 10c. Again, the reason is
presumably that the states of an electron in a singly and in a doubly occupied
site need to be sufficiently different, which will not happen if the two
orbitals are strongly mixed by $t'$.

It is interesting to examine the change in kinetic energy when carriers pair.
In Figure 11 we plot the difference between twice the kinetic energy of
a hole in the filled band and that of two holes in the filled band:
\beq
\Delta T=2<T>_{1hole}-<T>_{2 holes}
\eeq
with the kinetic energy operator given by Eq. (22). It can be seen by
comparison with Figure 10 that kinetic energy is always lowered
($\Delta T>0$) when carriers pair, i.e. in the regime where $U_{eff}$ is
attractive. The condition $\Delta T>0$ is necessary but not sufficient to
yield $U_{eff}<0$. This is because pairing is associated with a decrease in
kinetic energy and an increase in potential energy in this model.

\subsection{Comparison with conventional pairing}
In conventional models of superconductivity pairing arises from an effective
electron-electron attraction induced by coupling to a boson degree of 
freedom that does not differentiate between electrons and holes.
The resulting effective interaction is electron-hole symmetric, and in
such models pairing is driven by lowering of potential rather than
kinetic energy. We can describe such a scenario in our model by 
assuming negative values of the on-site interaction $U$, presumably resulting
from integrating out a boson. We also take a very large value of the interorbital spacing $\epsilon$ 
so as to approach a single band Hubbard model, and compute the
effective interaction between 2 holes (which is the same as between 2 electrons)
from Eq. (32). Not surprisingly, $U_{eff}$ for this conventional regime is
attractive (repulsive) when $U$ is attractive (repulsive).

It is interesting to compare the behavior of the  model in the
conventional regime (i.e. attractive Hubbard model regime) 
 with that in the regime discussed in the previous subsection where holes pair, which
we call 'dynamic Hubbard' regime. We compute the effective interaction
$U_{eff}$ versus $U'$ in the dynamic Hubbard regime and versus $U$ (positive and
negative) in the conventional regimen, and in figure 12
we plot various properties as a function of the resulting effective interaction $U_{eff}$ in both regimes.
Figure 12a shows the ratio of quasiparticle weights in the system with two holes
and one hole. In the conventional regime, $z_2/z_1$ is less than 1 both when the
effective interaction is attractive and when it is repulsive. In contrast, in the
dynamic Hubbard regime the quasiparticle weight for the case of two holes becomes
much larger than for a single hole when holes pair. Note however that $z_2$ can be
bigger than $z_1$ even when the effective interaction is still repulsive. 
In Fig. 12b we show the behavior of the effective hopping amplitude defined
from the difference in energy between the ground state and first excited state in 
the model. In the conventional regime, holes becomes heavier than single particles
($t_2/t_1<1$) when they pair, in contrast in the dynamic Hubbard regime paired holes
are much lighter than single holes when pairing occurs. However once again pairs
can be lighter than single holes even when the effective interaction is still repulsive.
This is of course a finite size effect because in an infinite system for low hole
concentration holes would be far from each other if not bound in a pair. Finally,
Fig. 12c compares the change in kinetic energy upon pairing. In the conventional regime
the kinetic energy increases upon pairing, so that the difference between twice the
single hole kinetic energy and the pair kinetic energy is negative, while the kinetic
energy decreases strongly upon pairing in the dynamic Hubbard regime.

These results illustrate the qualitative difference in the physics of pairing in these 
two different regimes. In the regime of dynamic Hubbard physics, pairing
is associated with undressing\cite{undressing}, i.e. increase in quasiparticle weight, decrease in
quasiparticle mass and lowering of kinetic energy. In the conventional regime, the 
physics of pairing is exactly opposite, pairing is associated with dressing, i.e. smaller
quasiparticle weight, larger quasiparticle mass and increase in kinetic energy for
the pair. In 
summary, upon pairing quasiparticles become more coherent and lighter
in the dynamic Hubbard regime
and more incoherent and heavier in the conventional regime.

Note that the attractive Hubbard model can describe conventional superconductivity both in the
short coherence length regime (large attractive $U$) and in the regime where the coherence
length is thousands of lattice spacings (small attractive $U$). The $qualitative$ contrast that we
make here between the physics of that model and the physics of the dynamic Hubbard model
applies to both regimes.

\section{Optical conductivity}

We calculate the optical conductivity given by Eq. (20). The total optical spectral weight in the model
is related to the expectation value of the kinetic energy operator as given
by Eq. (21). We will divide the frequency range into a low frequency range with
cutoff $\omega_m=2$, that defines the low frequency spectral weight $A_l$, and denote the
remaining spectral weight at higher frequencies by $A_h$. The low frequency spectral weight includes the
'intra-band' spectral weight; it also includes some low frequency absorption that is not 
intra-band when the lower band is close to full and $\epsilon $ is not too large.

Figure 13 shows the dependence of the integrated optical spectral weights on band filling,
for three sets of parameters. For Fig. 13(a), with $\epsilon=10$, the absorption is
approximately electron-hole symmetric; however even in this case with large $\epsilon$
it can be seen that for holes the intra-band low frequency absorption is somewhat lower and the 
high frequency absorption is somewhat higher than for electrons. As $\epsilon$ decreases
(Fig. 13(b)) and even more so when $U'$ also decreases (Fig. 13(c)), the intra-band
absorption becomes much smaller for holes than for electrons. Note also that for the
case of Fig. 13(b) where $U_{eff}$ is still repulsive between holes the low
frequency absorption for 2 holes is only slightly larger than for 1 hole; instead,
as $U'$ is decreased and $U_{eff}$ becomes attractive (Fig. 13(c)) the intraband
optical absorption for 2 holes becomes more than twice the intraband optical
absorption for 1 hole, because optical spectral weight is transfered from
high to low frequencies when pairing occurs.

Figure 14 compares the optical conductivity for the nearly empty and the nearly full band, for the
cases with $\epsilon=10$ and $\epsilon=2$. For the large $\epsilon$ case, the intraband
conductivity (per particle) is only slightly smaller for holes than for electrons.
Instead, for $\epsilon=2$ there is a dramatic difference in the optical
conductivity for electrons and holes: for holes, the intraband conductivity is very
small and most of the optical absorption occurs at higher frequencies.

Next we consider the behavior of the optical conductivity upon doping. For the case
of large $\epsilon$, it is similar for electrons and for holes, as seen in Figure 15:
the intraband conductivity per carrier decreases slightly with doping, and some
spectral weight is added at higher frequencies. For the case of small $\epsilon$, Figure 16,  the
behavior is similar for electrons but dramatically different for holes: 
in the latter case, there is a large increase in the low frequency spectral weight 
for the case of two holes, which is due to
the undressing of holes when they pair. Furthermore there is an overall shift of
the non-intraband spectral weight at higher frequencies to lower frequencies. Similar
behavior is found in other realizations of dynamic Hubbard models\cite{last}.
Such a transfer of optical spectral weight from high to low frequencies has been seen
in high $T_c$ cuprates upon hole doping and upon lowering the temperature below
the superconducting critical temperature\cite{uchida,marel,santander,basov,tanner}.

Finally we show the behavior of optical absorption in the regime of conventional
pairing. We choose a large value of the on-site attraction to illustrate the
behavior clearly, however the qualitative behavior persists for smaller attractive
interaction. The optical absorption is electron-hole symmetric as one would expect,
and most of the optical spectral weight is at low frequencies (intraband) in this
case for all band fillings. Comparing the case of 1 hole and 2 holes (or 1 electron
and 2 electrons) in Fig. 17 (b), it is seen that pairing is associated with a decrease
in the low frequency optical spectral weight, i.e. quasiparticles become more
dressed when they pair. This is in accordance with the behavior found for the
quasiparticle weight and effective hopping in figure 12, and qualitatively different
to the behavior in the dynamic Hubbard model regime.

\section{Relation with atomic physics and with real materials}

For any given atom one can relate the parameters in the site Hamiltonian
\beqn
H_i&=&Un_{i\uparrow}n_{i\downarrow}+U'n'_{i\uparrow}n'_{i\downarrow}
+\epsilon n'_i +V n_i n'_i \\ \nonumber
&&-t'\sum_\sigma (c_{i\sigma}^\dagger c'_{i\sigma} + h.c.)
\eeqn
to atomic quantities by comparison of properties obtained from it and properties
of the electronic states of the atom obtained from quantum chemical calculations.
As the simplest example we discuss here qualitatively the relation between
the Hamiltonian parameters and electrons in a hydrogenic ion with nuclear
charge $Z$ within the Hartree approximation.
The difference in energy between an electron in the $1s$ and $2s$ atomic orbitals
corresponds to the energy difference between the two single particle eigenstates
in Eq. (36), namely
\beq
\sqrt{\epsilon^2+4t^{'2}}\sim 13.6\times Z^2 \times \frac{3}{4}
\eeq
in $eV$ units here and in what follows. For small $t'$ we have approximately
\beq
\epsilon\sim 10.2 Z^2
\eeq
We will assume $t'$ small in what follows so that the strong coupling analysis
is applicable. The repulsion $U$ in the lower orbital corresponds to
the repulsion of 2 electrons in the $1s$ orbital
\beq
U=17Z
\eeq
In the Hartree approximation the single electron orbital with wavefunction
$\varphi \propto e^{-Zr}$ expands upon double occupancy to wavefunction
$\bar{\varphi} \propto e^{-\bar{Z}r}$, with
\beq
\bar{Z}=Z-\frac{5}{16}
\eeq
We identify the Coulomb repulsion in the upper orbital, $U'$, as the
repulsion of two electrons in the Hartree expanded orbital, i.e.
\beq
U'=17\bar{Z}=U-5.31
\eeq
We can estimate the Coulomb repulsion between electrons in the two different
orbitals, $V$, by calculating the Coulomb integral for one electron in the 
$1s$ orbital and another in the expanded Hartree orbital. This yields (in eV)
\beq
V=27.2 Z \bar{Z} \frac{Z^2+3Z\bar{Z}+\bar{Z}^2}{(Z+\bar{Z})^3}
\eeq
Finally we can estimate the intra-atomic hopping $t'$ from the overlap
matrix element of the single electron wave function in the doubly and
singly occupied sites. In the model, that is approximately
\beq
S=<\tilde{\downarrow}|c'_\uparrow|\tilde{\uparrow\downarrow}>
=t'(\frac{1}{\epsilon}+\frac{1}{V-U'-\epsilon})
\eeq
and in the Hartree atom it is
\beq
S=<\varphi|\bar{\varphi}>=\frac{(Z\bar{Z})^{3/2}}{((Z+\bar{Z})/2)^3}
\eeq

It can be seen that in the atom $U>V>U'$ for all $Z$. As $Z$ decreases, all Coulomb
repulsions decrease, as well as $\epsilon$ and the overlap $S$. This is the
regime favorable for pairing in this model. The Hartree calculation is of course
very approximate, and in particular it overestimates the overlap matrix element
$S$. Nevertheless it illustrates the basic trend. For orbitals higher than the
$1s$ the energy levels become closer in energy and the effects discussed in
this paper should become stronger.

In summary, the atomic charge $Z$, with $Z-2$ the charge of the ion when the relevant band
is full, is the key atomic parameter. For small $Z$ the parameters in the
Hamiltonian studied in this paper move towards the regime of interest,
namely small Coulomb repulsion $U'$, small interband separation $\epsilon$ and 
small overlap matrix element $S$. In that regime electron-hole asymmetry in the band
becomes dominant, holes become heavily dressed in the normal state and they
strongly undress when they pair.

For high $T_c$ cuprates the relevant band of interest is one formed by overlapping
planar oxygen $p\pi$ orbitals in the $CuO_2$ planes\cite{twoband}. Since in the undoped system (no holes)
the ion is $O^=$, $Z=0$ in this case. For $MgB_2$, the relevant band is formed
by overlapping boron $p_{xy}$ orbitals in the $B^-$ planes\cite{mgb2}, and $Z=1$. The fact
that the planes are negatively charged in both cases ($Z<2$) favors the
physics discussed here, with the effects stronger for the cuprates due to the
smaller $Z$. Even stronger hole dressing and higher $T_c$'s would be expected
in a structure with even smaller $Z$, for example if one managed to make
a material with $N^{\equiv}$ planes doped with some holes ($Z=-1$).

The material $LiBC$ has been recently proposed as a candidate for 
high temperature superconductivity when hole-doped, by analogy with
$MgB_2$, within electron-phonon theory\cite{pickett}. Because the $(BC)^-$ planes in
that material would be less negatively charged than the $(B_2)^=$ in $MgB_2$, i.e. 
effectively $Z=1.5$ instead of $Z=1$, we expect this not to be a modification of
$MgB_2$ conducive to higher $T_c$'s  within the physics discussed here.
If such material was found to have a $T_c$ larger than $MgB_2$, as predicted\cite{pickett},
 it would directly
contradict the assumptions of this paper and prove the inapplicability of the
concepts discussed here to real materials.

\section{Discussion}

Electrons in solids interact with each other with an interaction strength
($e^2=14.4 eV A$) that is of the same magnitude as the interaction strength of
electrons with ions. It was recognized from the beginnings of solid state
physics that Bloch's approach of prioritizing the electron-ion interaction
over the electron-electron interaction was an ad-hoc assumption that could
certainly not be rigurously justified. Even though Landau's Fermi liquid theory
with the concept of a quasiparticle provides an explanation for the fact that
many properties of solids look amazingly 'independent-electron-like', the
fundamental role of electron-electron interactions in solids is still not
well understood.

This paper is part of a continuing effort to understand the role of electronic
correlation in energy bands. We argue that a key fact that has been ignored in
previous treatments of the problem is the dependence of quasiparticle weight on
band filling and the fundamental role of electron-hole asymmetry. In this paper we
studied a 'minimal model' that incorporates these key features. We believe that this physics
is part of the physics of all electronic energy bands: that
quasiparticle weights at the Fermi level when the band filling $n$ ($0<n<2$) is below and above
the half-filled band ($n=1$) are related by
\beq
z(n)>z(2-n)
\eeq
with $n<1$, i.e. that holes are $always$ more dressed than electrons. How different the
quasiparticle weights in the lower and upper halfs of the band are determines
how important the new physics originating in this effect is. This in turn depends
on the ionic charge $Z$ ($Z-2=$ionic charge when the band is full) with the
strongest effects occurring for small $Z$. Because quite generally in an atom
the intra-orbital Coulomb repulsion is linear
in $Z$ (e.g. Eq. (39)) and the energy level spacing is quadratic in $Z$ (e.g. Eq. (38))
the effects discussed here will become unimportant for sufficiently large $Z$.

We have called the models describing this physics 'dynamic Hubbard models'. In these
models, unlike the case of the conventional Hubbard model, the strength of the on-site
repulsion $U$ becomes a dynamical variable that can take more than one value depending on
the state of the two electrons in the atom. Here this dynamics is incorporated by
having two electronic orbitals per site; in other work we have described this dynamics
with a single electronic orbital per site and an auxiliary boson degree of 
freedom\cite{qmc,last,dynh,hole}.
While the model discussed here is more realistic and closer to the physics of real
atoms, the models with auxiliary boson degrees of freedom are simpler to treat
theoreticaly and thus may yield useful insight into the fundamental physics
of this class of models. From the results in this paper and in previous work we
believe that dynamic Hubbard models with only electronic degrees of freedom and those 
with auxiliary boson degrees of freedom share the same fundamental physics.

We have studied the two-orbital model by exact diagonalization of a small cluster. It
should be possible to study larger clusters with more computing power and more
sophisticated numerical techniques such as Lanczos diagonalization, density-matrix
renormalization group and quantum Monte Carlo. We believe that the qualitative physics
found here is likely to exist in larger systems.

The calculations in this paper yield the properties of interacting electrons in a model 
Hamiltonian for
the entire range of band fillings from empty to full, without uncontrolled 
approximations. Before this work such studies had only been performed for simpler
models such as the single band conventional Hubbard model, which as we have argued
lacks some essential physics. The results found here should qualitatively apply to 
all electronic energy bands in solids.

The results found here corroborate some of our earlier findings concerning the
importance of electron-hole asymmetry\cite{undressing} and
display clearly the interpolation between the conventional understanding of 
electronic correlations in energy bands and the physics stressed in the theory of
hole superconductivity. 
In the conventional understanding electrons and holes are
similar, quasiparticles are undressed when the band is almost empty and almost full,
and the dressing and importance of electron-electron interactions increases as one
approaches the half-filled band from either side. Instead, in the theory of hole
superconductivity in its simplest interpretation the dressing of a quasiparticle increases
monotonically as the band filling increases from the empty to the full band. 
As we have seen in this paper, the actual situation is always in-between these two
limiting descriptions, with the relationship Eq. (45) holding in all cases. 

The essential difference between conventional (static) and dynamic Hubbard models 
concerning 'intra-band' physics is that the state of a given electron is the same
in the singly and doubly occupied atom in the static Hubbard model, while it is
different in the dynamic Hubbard model. Through this modification of the state
the intraband bare particles, which were strongly interacting with repulsion $U$,
become weakly interacting quasiparticles with interaction $U'$. This occurs at the
level of a single site, and is expressed by the relation between bare particle
operators $c_{i\sigma}$ and quasiparticle operators $\tilde{c}_{i\sigma}$
\beq
c_{i\sigma}=[1+(S-1)\tilde{n}_{i,-\sigma}]\tilde{c}_{i\sigma}
\eeq
The quasiparticle dynamics is described by the kinetic energy Eq. (30) and the
local repulsion $U'$, and their weight is further modified by the
weak interactions in the quasiparticle band. The quasiparticle weight 
at the Fermi energy can be approximately written as
\beq
z(n)=[1+(S-1)\frac{n}{2}]^2 z_{ib}(n)
\eeq
with $0\leq n \leq 2$ the band filling, and the 'intraband' quasiparticle weight
$z_{ib}$ defined by
\beq
z_{ib}=|<0_{N-1}|\tilde{c}_{k_F\sigma}|0_N>|^2
\eeq
calculated using the ground states of the Hamiltonian for the intraband
quasiparticles (Eq. (30) plus weak on-site repulsion). In particular 
$z_{ib}(n\rightarrow 0)=z_{ib}(n\rightarrow 2)=1$ and
is smallest near the half-filled band, as in the conventional 
Hubbard model. The factor multiplying $z_{ib}$ in Eq. (47) isolates the main
effect of electron-hole asymmetry. However even $z_{ib}$ will exhibit some additional
electron-hole asymmetry (of the same sign) due to the 
dependence of the effective bandwidth on filling Eq. (33): the residual intraband interactions
will more strongly dress the quasiparticles in the upper half of the band where the
effective bandwidth is smaller.

For materials with large ionic charge $Z$, $S$ will be close to $1$,
 electrons and holes will be
very similar and the dominant dressing will occur near the half-filled band. Instead,
for materials with small $Z$, $S$ will be much smaller than $1$,
 the physics of hole
superconductivity will dominate, holes will be highly dressed near the full band
and strongly undress as the local hole concentration increases.
We propose that the physics of high temperature superconductivity in solids is described 
by the latter regime. As the parameters become less extreme with increasing ionic
charge $Z$ the dressing of holes in the normal state becomes less extreme, the 
undressing effect of pairing becomes less apparent, the coherence length of the
Cooper pairs increases and one moves towards the regime of 'conventional' superconductivity\cite{parks}.

\acknowledgements
The author is grateful to Fred Driscoll for providing the computer facilities where 
this work was performed. This work was NOT supported by the National Science Foundation.

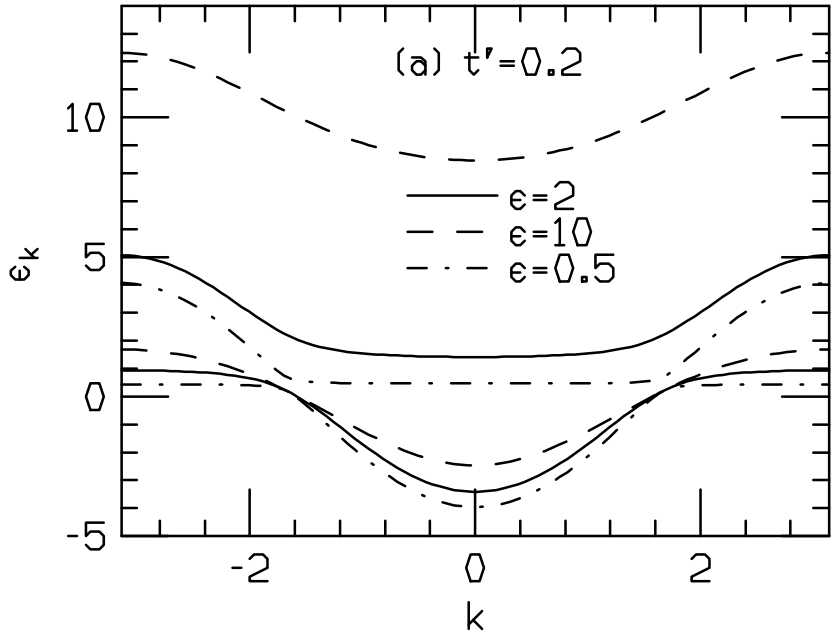
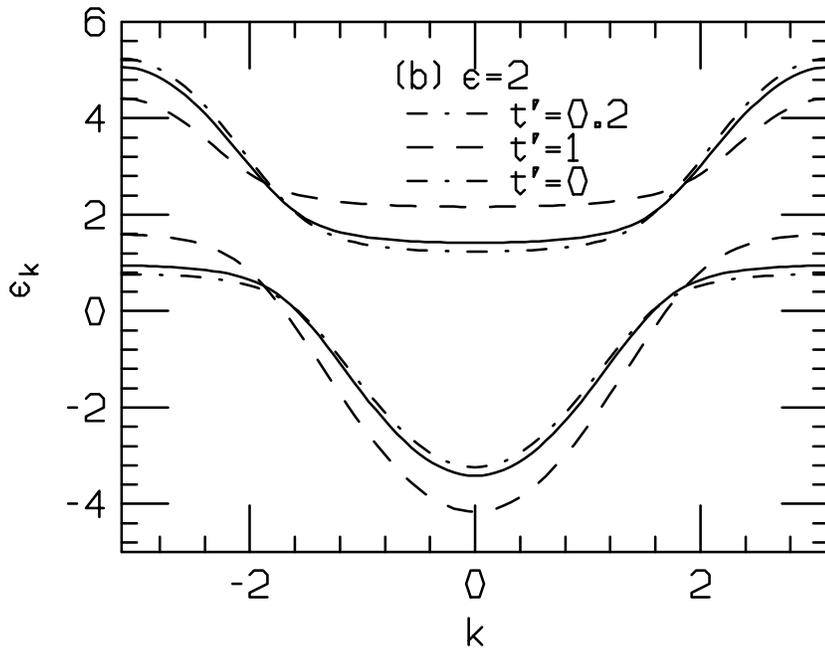
\begin{figure}
\caption { Non-interacting band structure for the model under consideration
for various parameters. (a) As $\epsilon$ increases, the gap between lower
and upper band increases. (b) As $t'$ increases the bandwidth of the lower
band increases and that of the upper band decreases.
}
\label{Fig. 1}
\end{figure}
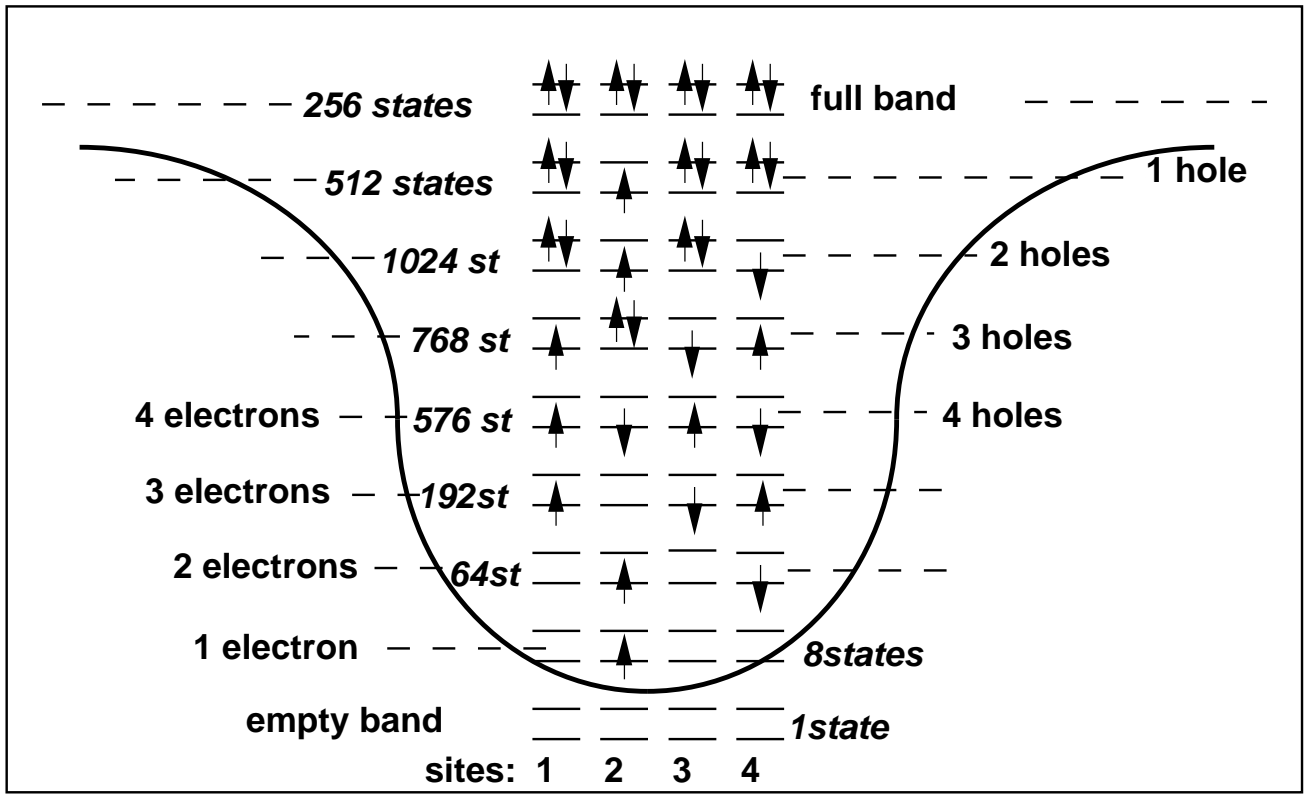
\begin{figure}
\caption {Schematic depiction of states in 4-site cluster for different band
filling. Because of the large $U$ in the lower orbital, two electrons at the
same site will predominantly occupy the higher orbital.
The number of states in the Hilbert space for each filling is also
given. 
}
\label{Fig. 2}
\end{figure}
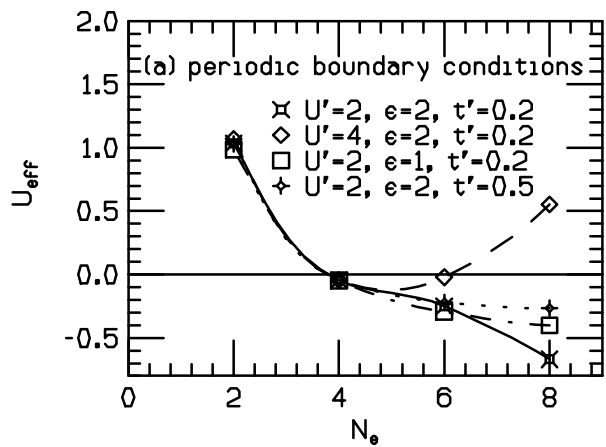
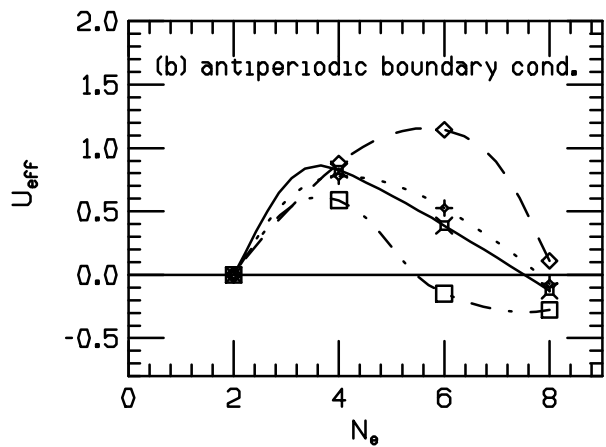
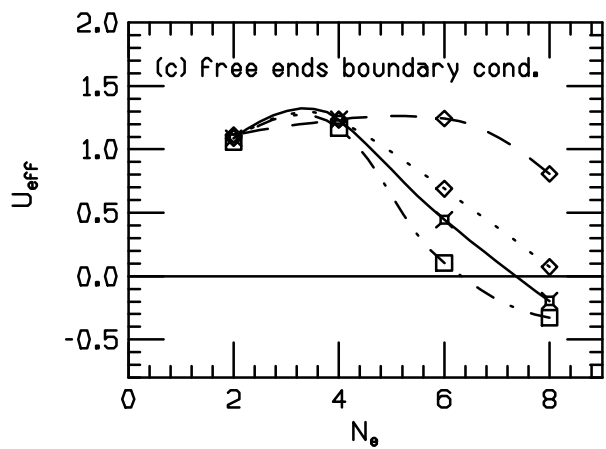
\begin{figure}
\caption {Effective interaction Eq. (32) versus band filling for different 
boundary conditions, for four sets of parameters. $t=1$ here and
in the following figures. For all cases here $U=10$ and $V=6$.
The lines through the data 
are a guide to the eye. Solid line: $U'=2, \epsilon=2, t'=0.2$; dashed line: 
$U'=4, \epsilon=2, t'=0.2$; dash-dotted line:  $U'=2, \epsilon=1, t'=0.2$; 
dotted line: $U'=2, \epsilon=2, t'=0.5$.
}
\label{Fig. 3}
\end{figure}
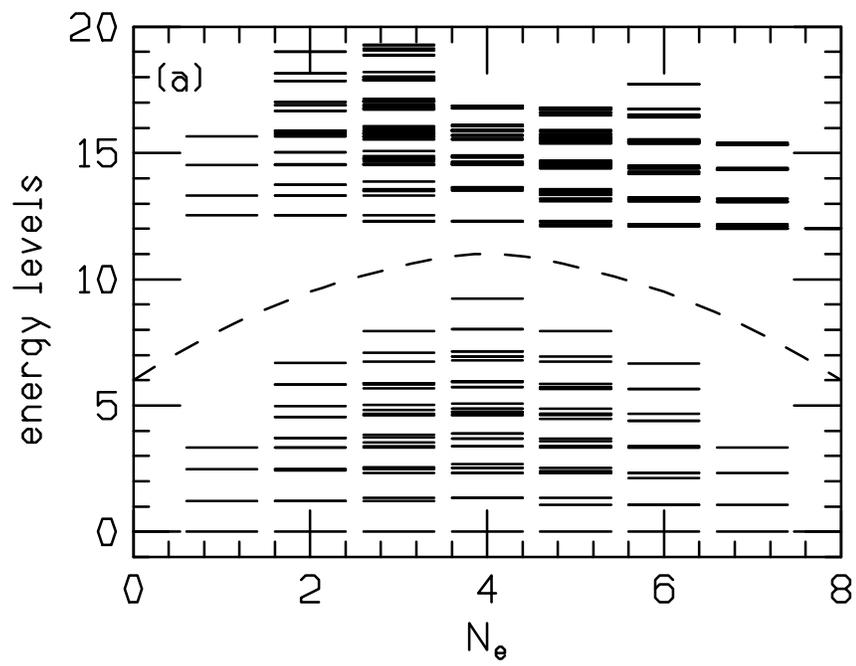
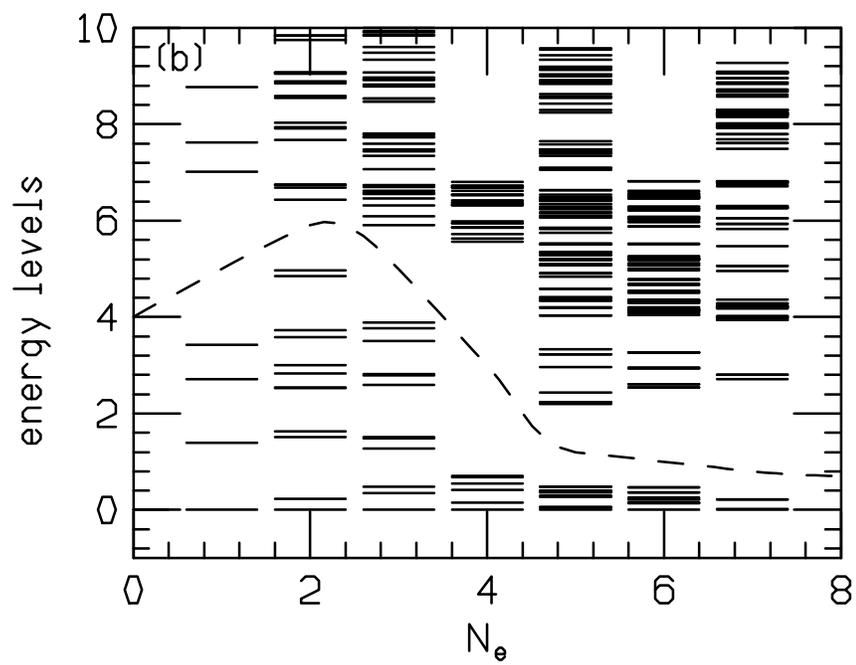
\begin{figure}
\caption {Energy eigenvalues for the different band fillings. Parameters are:
(a) $\epsilon=12$, $t'=0.2$, $U=V=U'=0$;
(b) $\epsilon=6$, $t'=0.2$, $U=20$, $V=12$, $U'=2$. The dashed lines show
the separation between the 'intra-band' states described approximately
by the Hamiltonian Eq. (3) below it, and the rest of the states in the Hilbert space. 
For other less extreme parameters no clear separation is seen. 
Note that in the interacting case (b) the spectrum is strongly 
electron-hole asymmetric, with the lower band bandwidth for holes ($N_e>4$) much smaller than
for electrons ($N_e<4$).
}
\label{Fig. 4}
\end{figure}
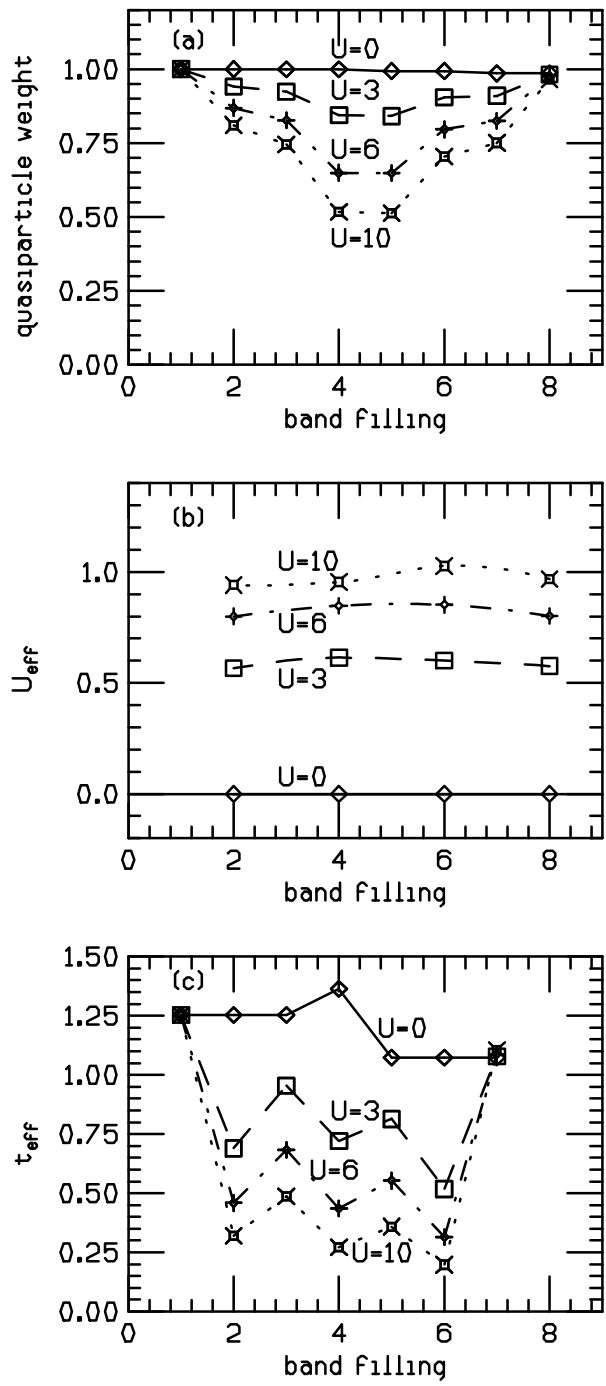
\begin{figure}
\caption {The parameters used are $\epsilon=10$, $t'=0.2$ 
and the following interactions: solid lines: $U=V=U'=0$; 
dashed lines: $U=3, V=2, U'=1$; 
dash-dotted lines: $U=6, V=4, U'=3$; 
dotted lines: $U=10, V=6, U'=5$. Plotted versus bandfilling
$N_e$ are (a) quasiparticle weight at the Fermi energy, Eq. (17);
(b) effective interaction Eq. (32); (c) effective hopping defined
as the energy gap between the ground state and the first excited state.
Note that even for this case of large $\epsilon$ there is a small
electron-hole asymmetry, with the quasiparticle weight and effective
hopping being slightly smaller for holes than for electrons.
}
\label{Fig. 5}
\end{figure}
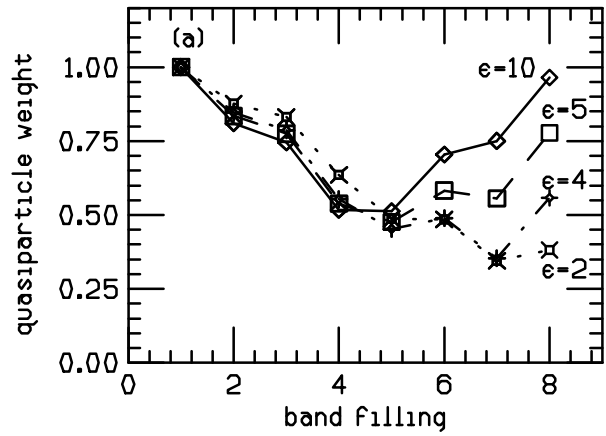
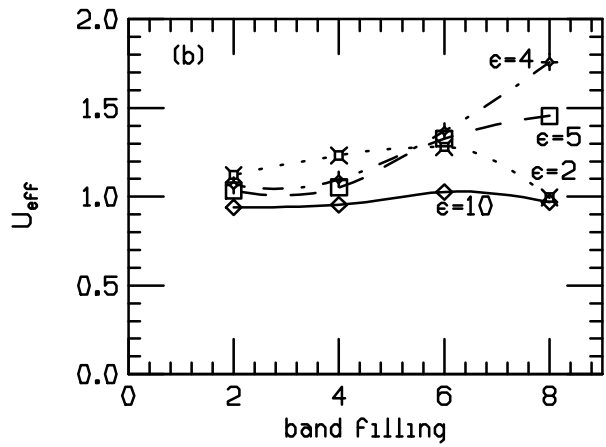
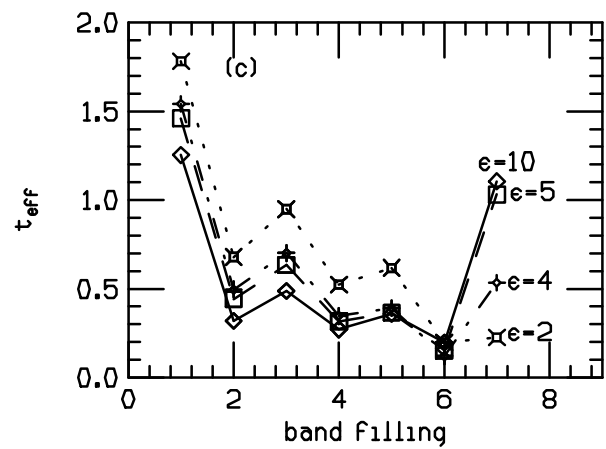
\begin{figure}
\caption {Same as figure 5 for $t'=0.2$, $U=10, V=6, U'=5$ and
$\epsilon$ values given in the figures. As $\epsilon$ decreases
holes become more incoherent and heavier, i.e. smaller $z$ and $t_{eff}$.
}
\label{Fig. 6}
\end{figure}
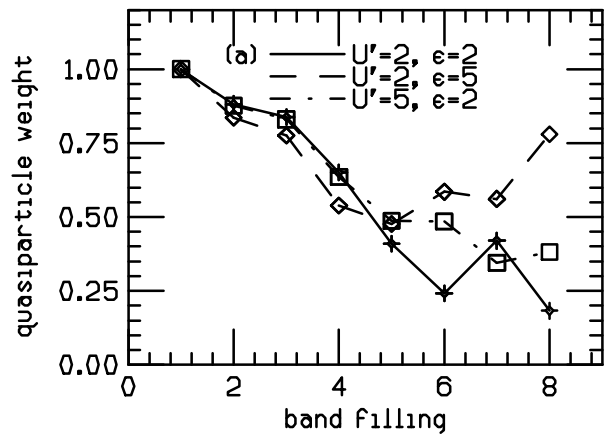
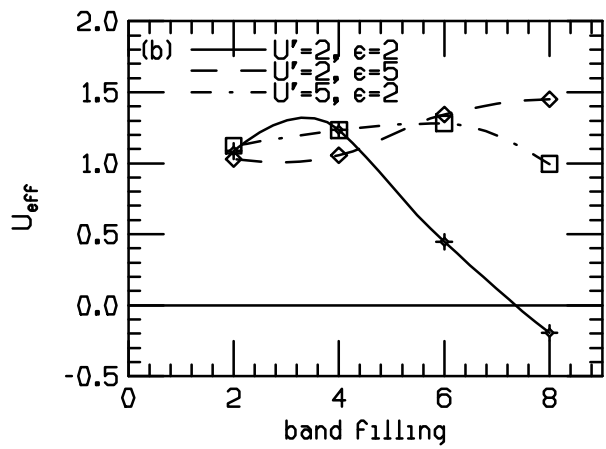
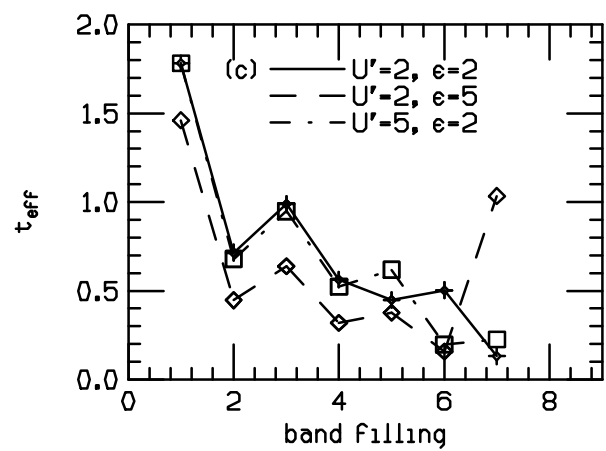
\begin{figure}
\caption {Same as figure 5 for $t'=0.2$, $U=10, V=6$ and
values of $\epsilon$ and $U'$ given in the figure. For small
$\epsilon$ and $U'$ (solid lines) the effective interaction is attractive for
holes; in that case, the quasiparticle weight at the Fermi energy
for the band filled with 2 holes is larger than
for the band with 1 hole (a), and the effective hopping is larger for the band
with 2 holes than for the band with 1 hole.
}
\label{Fig. 7}
\end{figure}
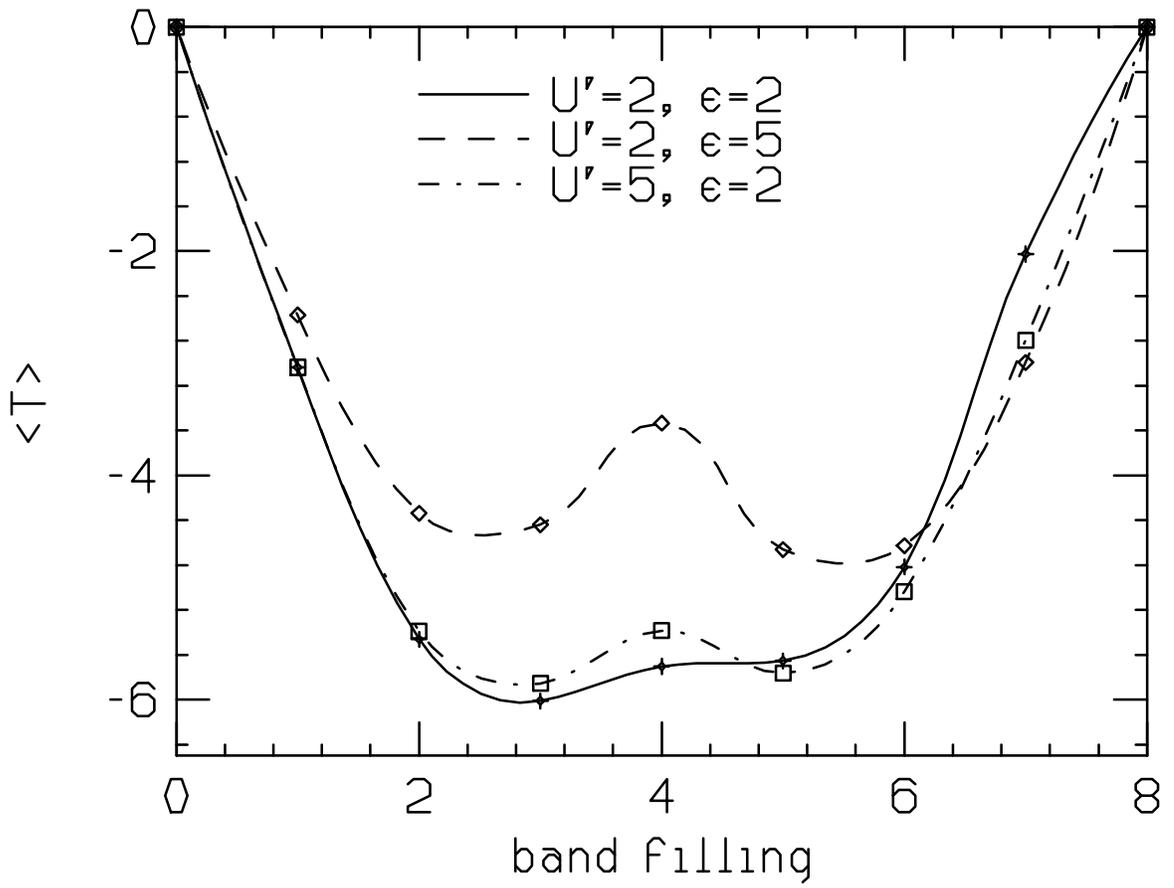
\begin{figure}
\caption {Expectation value of the kinetic operator Eq. (22) for the same
parameters as in Fig. 7. Note that for the case of attractive effective
interaction (full line) the kinetic energy for two holes is lower than
twice the kinetic energy for one hole, while in the cases of repulsive
effective interaction it is higher.
}
\label{Fig. 8}
\end{figure}
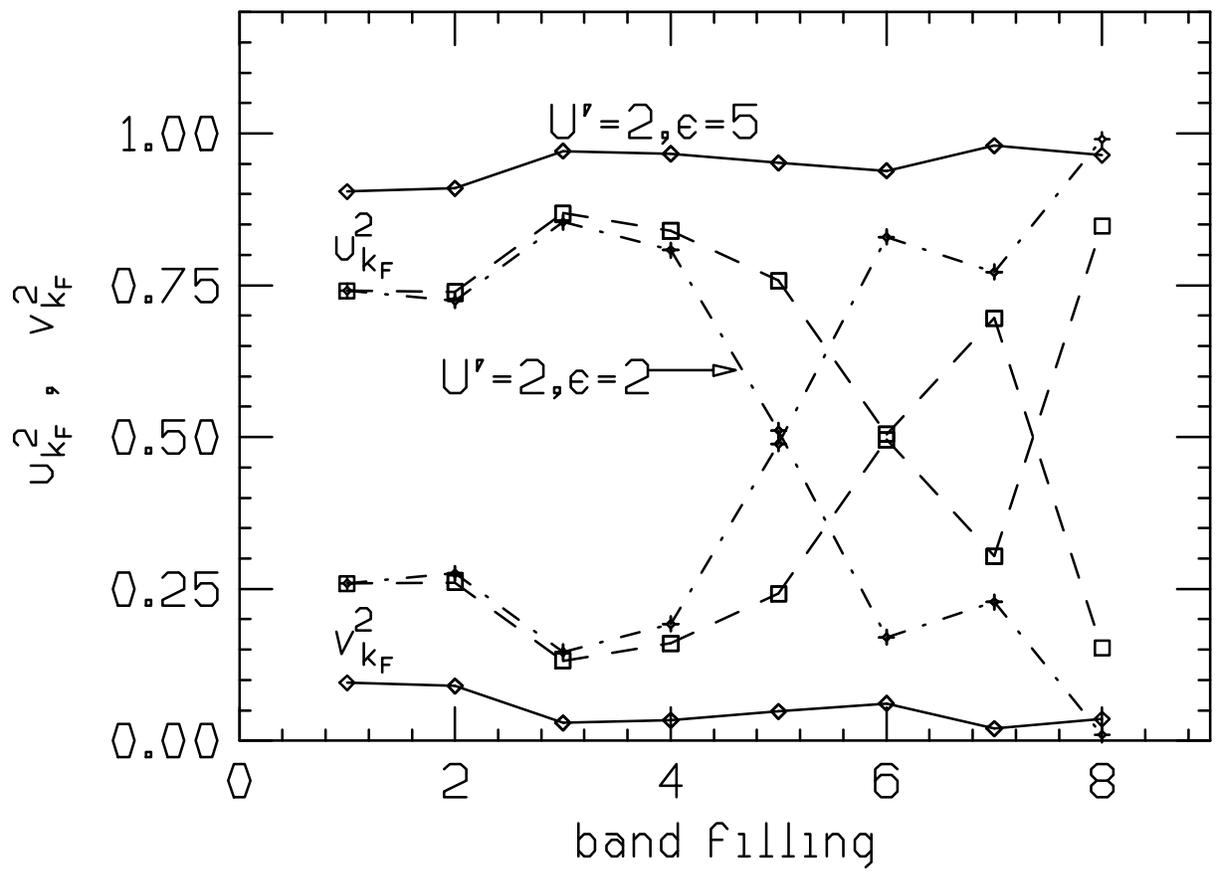
\begin{figure}
\caption {Composition of the quasiparticles, from Eq. (18), for the parameters
of Fig. 7. 
$u_{k_F}$ and $v_{k_F}$ give the amplitude of the quasiparticle at the
Fermi energy in the lower and upper atomic orbital. As the band
filling increases $u_{k_F}$ decreases and $v_{k_F}$ increases, with the
changes being largest for the case of attractive effective interaction
}
\label{Fig. 9}
\end{figure}
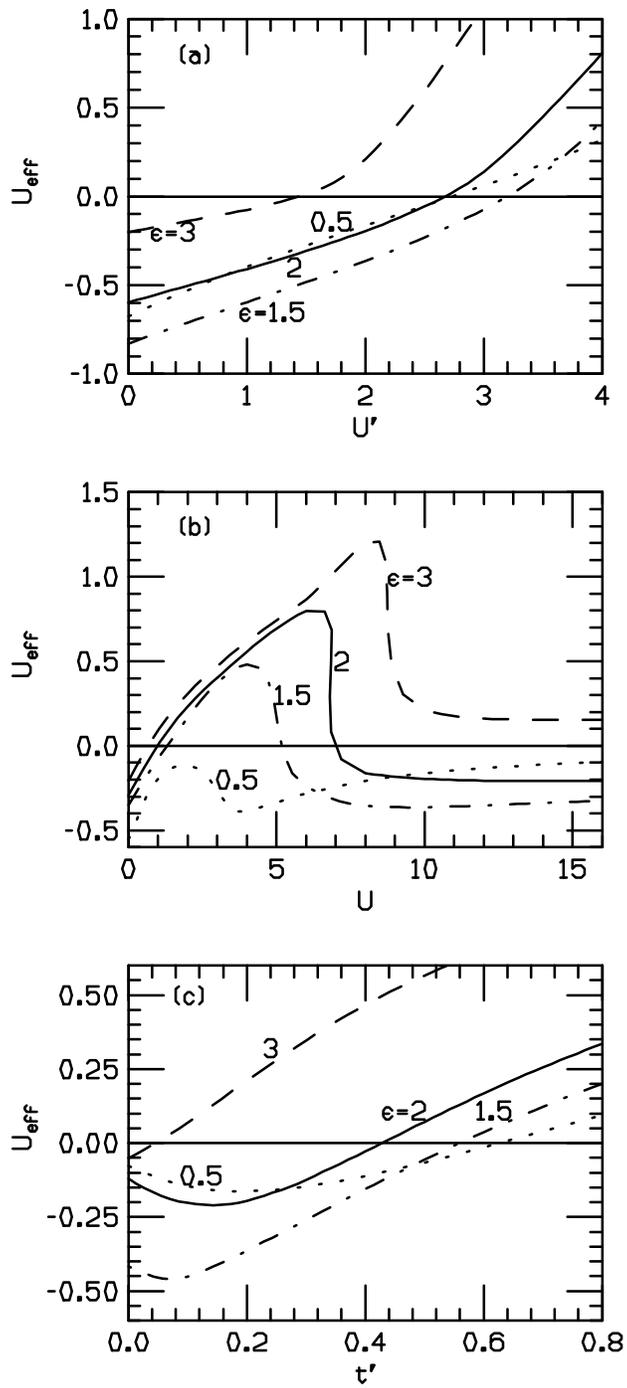
\begin{figure}
\caption {Dependence of effective interaction for two holes in the
filled band on Hamiltonian parameters. The values of 
$\epsilon$ are given in the figures. (a) Versus $U'$, with $U=10, V=6, t'=0.2$;  
(b) versus $U$, with $U'=2, V=6, t'=0.2$; 
(c) versus $t'$, with $U=10, V=6, U'=2$.
}
\label{Fig. 10}
\end{figure}
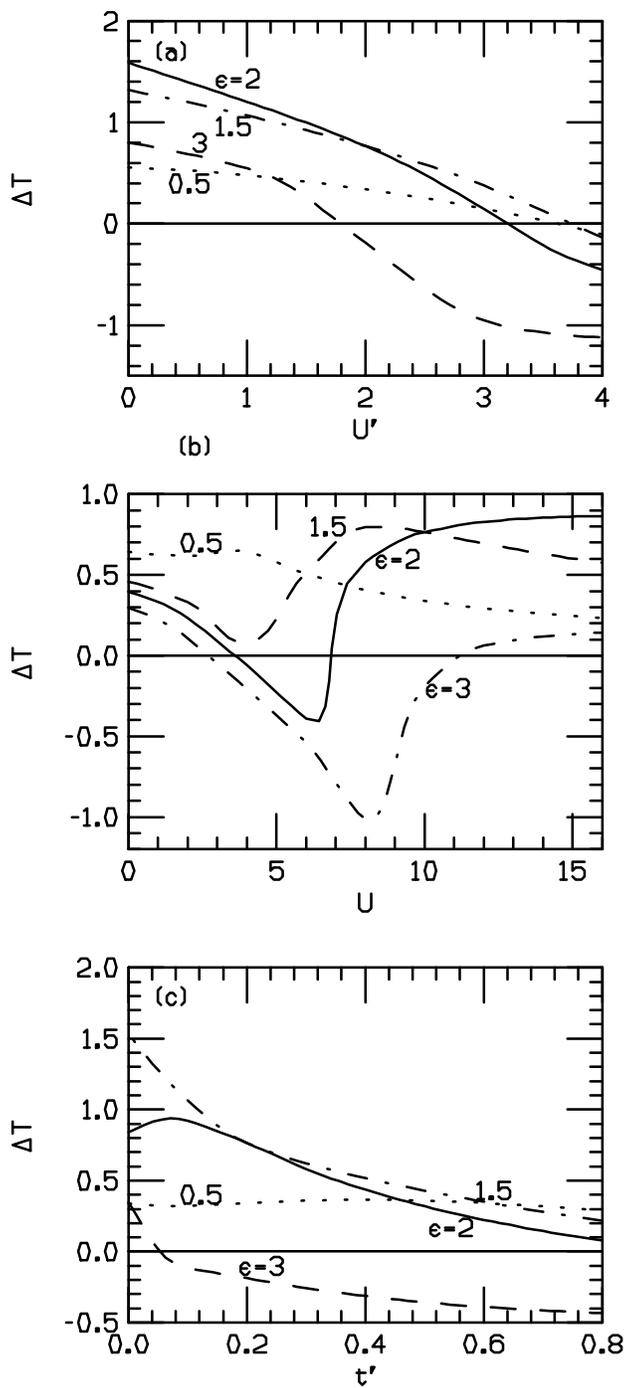
\begin{figure}
\caption {Dependence of kinetic energy difference between single holes
and pair of holes, Eq. (35), on Hamiltonian parameters,  for the same
cases as Fig. 9.
}
\label{Fig. 11}
\end{figure}
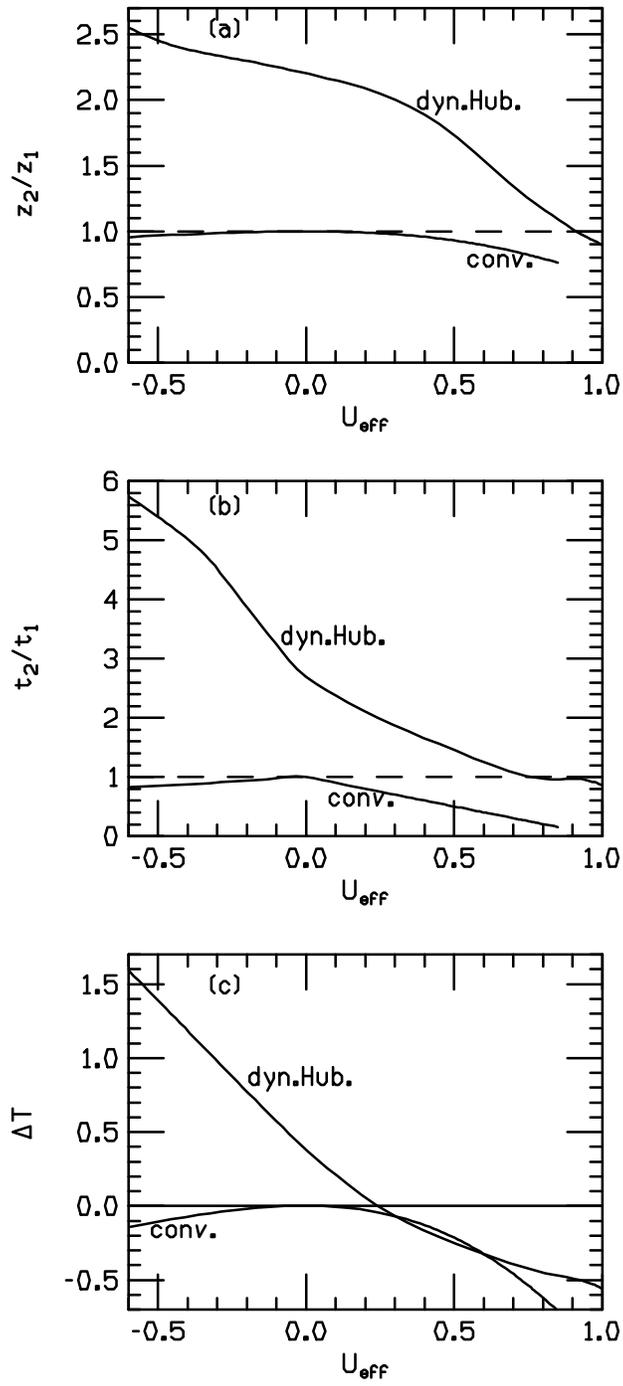
\begin{figure}
\caption {Comparison of behavior of model in conventional (electron-hole symmetric)
regime and regime of dynamic Hubbard physics. Parameters used for
dynamic Hubbard regime are $U=10, V=6,  \epsilon=2, t'=0.2$
and $U'$ ranging from 0 to 6; for conventional regime,
$\epsilon=100, t'=0.2, V=6, U'=5$ and $U$ ranging from -1.5 to 8. In both cases
$U_{eff}$ is calculated from Eq. (32) and the results are plotted versus
$U_{eff}$. (a) Ratio of quasiparticle weights for two holes and for one hole;
(b) ratio of hopping amplitudes for two holes and one hole; 
(c) difference in kinetic energy of single holes and paired holes, Eq. (35).
}
\label{Fig. 12}
\end{figure}
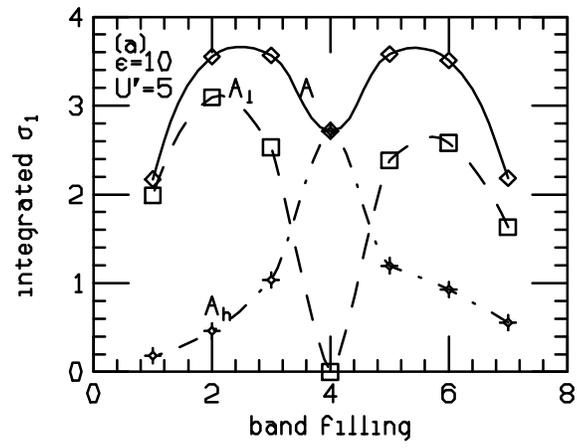
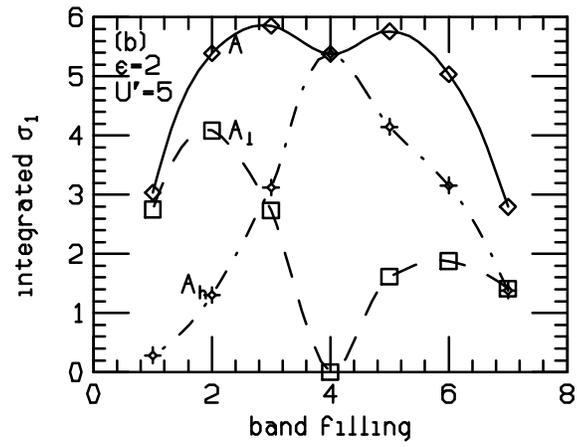
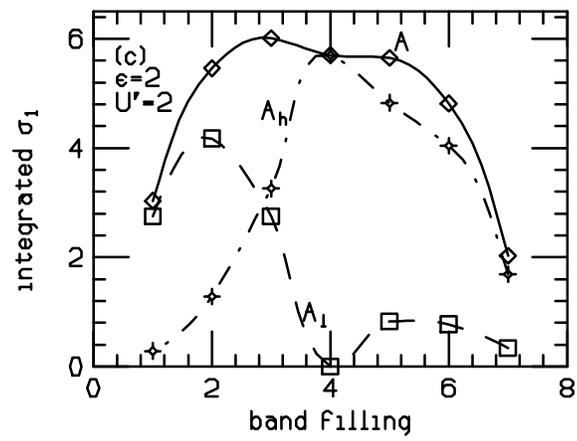
\begin{figure}
\caption {
Integrals of optical conductivity. $A_l$ denotes the low frequency integral
Eq. (23), with cutoff $\omega_m=2$; $A_h$ denotes the high frequency
 optical spectral weight for $\omega > \omega_m$, and
$A=A_l+A_h$ the total optical spectral weight Eq. (21).
Parameters are $U=10, V=6, t'=0.2$ and values of $U'$ and $\epsilon$ given
in the figures (same parameters as Figs. 7-9).
}
\label{Fig. 13}
\end{figure}
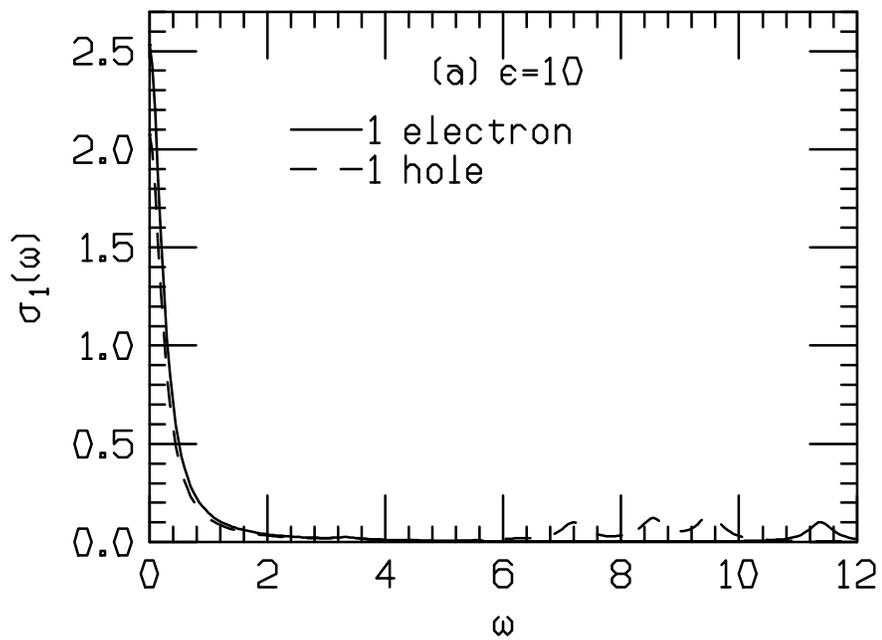
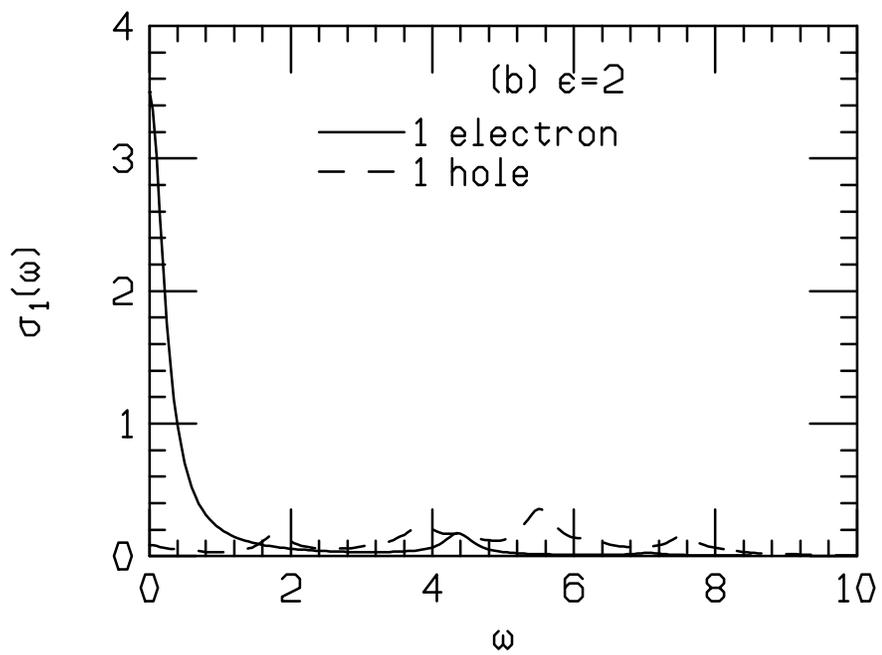
\begin{figure}
\caption {Comparison of optical conductivity for one electron and
one hole for $t'=0.2$ and values of $\epsilon$ given in the figure.
Here and in the following figures of optical conductivity, the $\delta$-functions
in Eq. (20) are broadened to Lorentzians with width $\Gamma=0.25$. 
The lowest frequency $\delta-$function at the 'Drude precursor' frequency\protect\cite{hanke}
is shifted to $\omega=0$ and represented by a Drude form (semi-Lorentzian).
Note that for large $\epsilon$ the conductivities for electrons and holes are very similar, for small
$\epsilon$ the conductivity for holes is very small and in particular
the intra-band conductivity is much smaller than for the case of large $\epsilon$.
}
\label{Fig. 14}
\end{figure}
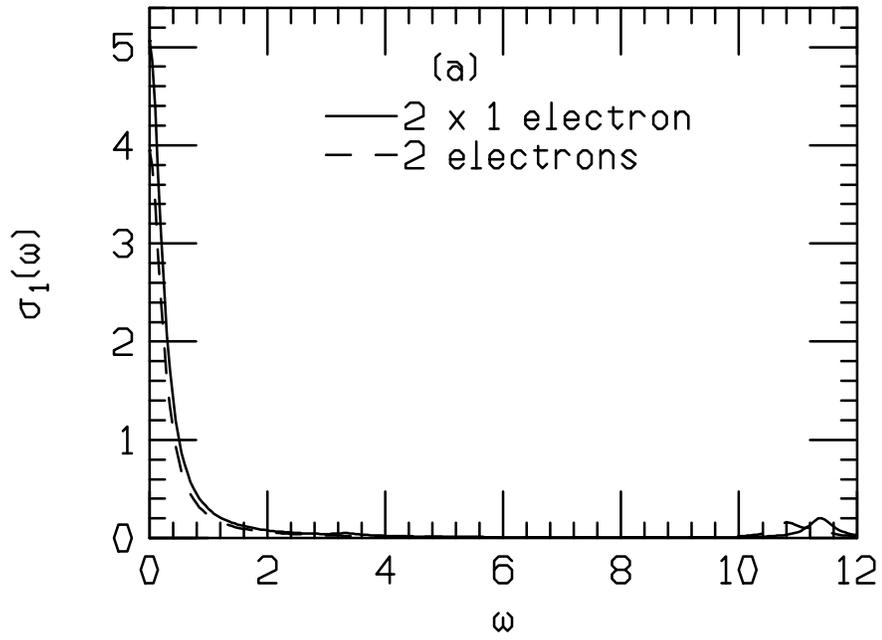
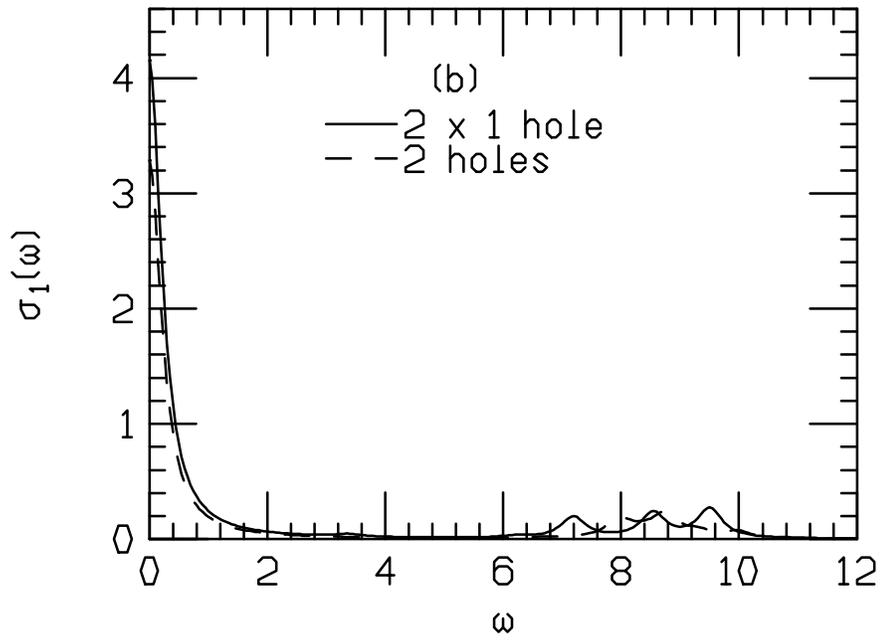
\begin{figure}
\caption {Comparison of optical conductivity for 1 carrier and 2 carriers in
the band. The optical conductivity for one carrier is multiplied by a factor 2
to make it comparable to the optical conductivity for two carriers.
(a) electrons, (b) holes. Here and in the next figure  $U=10, V=6, t'=0.2$.
Here, $\epsilon=10, U'=5$. Note that the optical conductivity is
similar for 1 and 2 carriers (normalized to number of carriers) both 
for electrons and for holes, with the low frequency absorption (per carrier) being slightly
smaller when the number of carriers is larger.
}
\label{Fig. 15}
\end{figure}
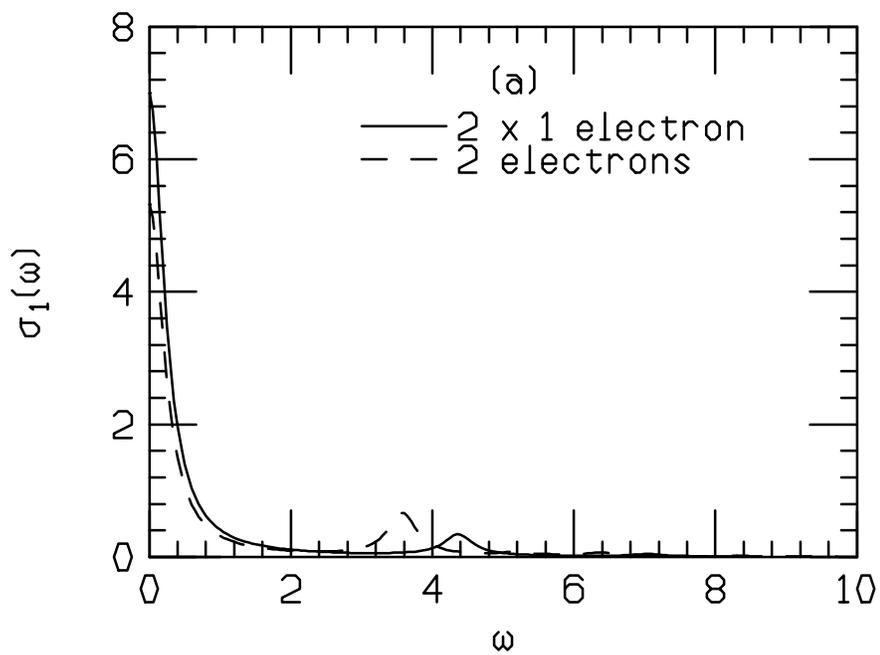
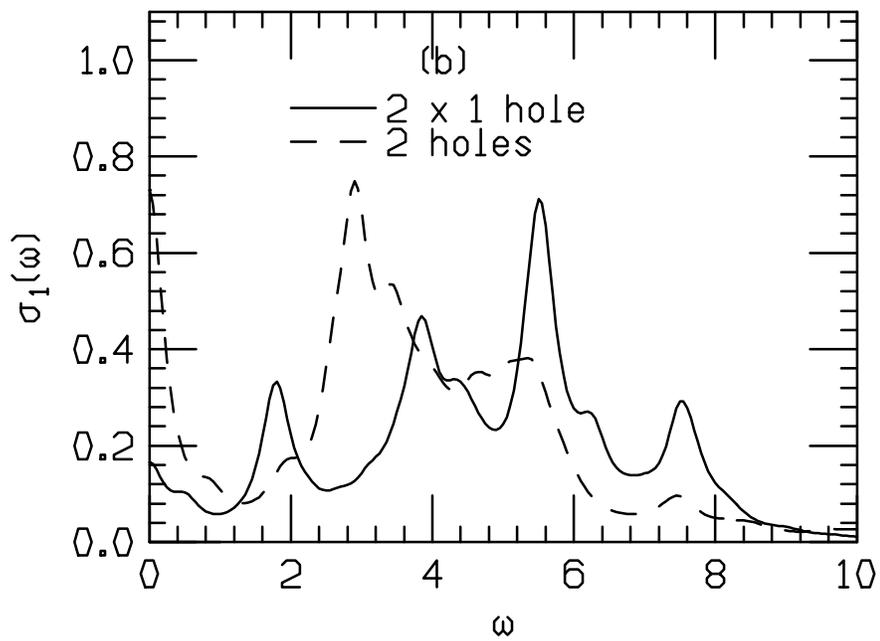
\begin{figure}
\caption {Same as Fig. 15 for parameters $\epsilon=2, U'=2$.
For the case of electrons the results are similar to Fig. 15, for the case
of holes they are very different: for the case of 2 holes there is a large
increase in low frequency absorption and an overall shift in optical absorption to 
lower frequencies.
}
\label{Fig. 16}
\end{figure}
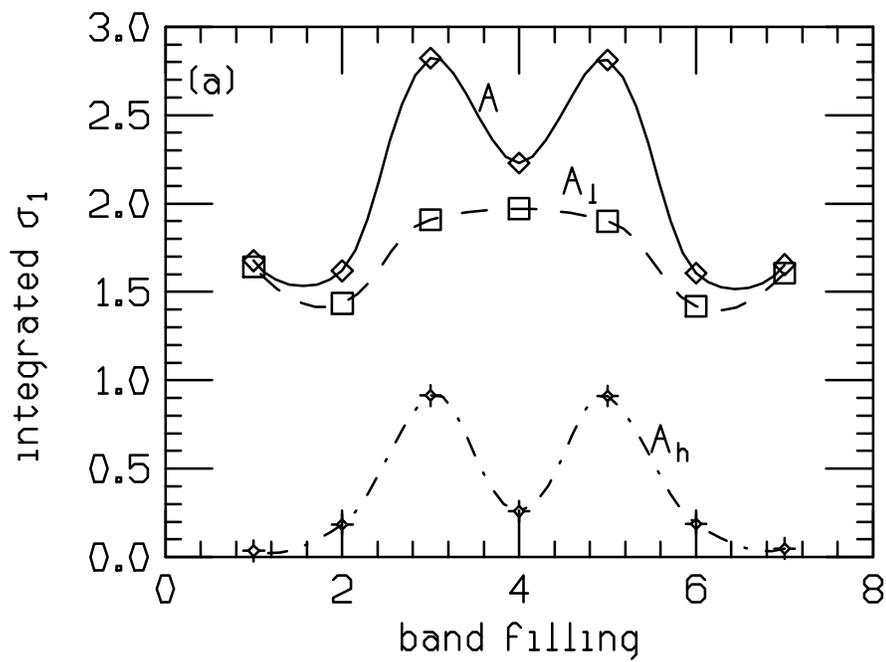
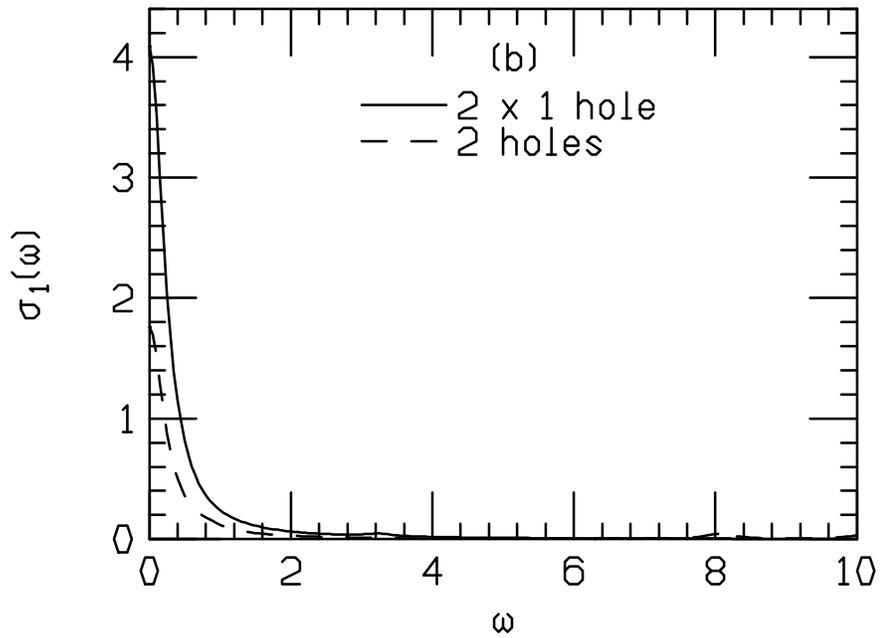
\begin{figure}
\caption {Optical absorption for parameters in the 'conventional' regime, given
in the caption of Fig. 12, with strong on-site attractive interaction $U=-8$,
giving rise to effective attraction $U_{eff}=-5.6$. (a) Integrated optical absorption,
same cutoff as in Fig. 13. Note that the low frequency absorption does not
increase as the number of carriers increases from 1 to 2.
(b) Optical conductivity for 1 and 2 holes (essentially the same as for 1 and 2 electrons
for these parameters). Note that upon pairing the low frequency absorption
decreases strongly as paired carriers are more highly dressed.
}
\label{Fig. 17}
\end{figure}


\begin{references}
\bibitem{slater} J.C. Slater, ``Quantum Theory of Atomic Structure'', Mc Graw Hill, New 
York, 1960.
\bibitem{hubbard} J. Hubbard, Proc.Roy.Soc. London A276, 238 (1963).
\bibitem{atom} J.E. Hirsch, Phys.Rev.B {\bf 65}, 184502 (2002) and references therein.
\bibitem{deghub} L.M. Roth, Phys.Rev. {\bf 149}, 306 (1966); K.I. Kugel and 
D.I. Khomskii, Sov.Phys. JETP {\bf 37}, 725 (1973); M.Cyrot and C. Lyon Caen, J. Phys.
(Paris) {\bf 36}, 253 (1975); W. Gill and D.J. Scalapino, Phys.Rev.B {\bf 35}, 215 (1987);
K. Kuei and R.T. Scalettar, Phys.Rev.B {\bf 55}, 14968 (1997).
\bibitem{falkim} L.M. Falicov and J.C. Kimball, Phys. Rev. Lett. {\bf 22}, 997 (1969);
Q.Si, G. Kotliar and A. Georges, Phys. Rev. B {\bf 46}, 1261 (1992);
P. Farkasovsky, Phys. Rev. B {\bf 51}, 1507 (1995);
T. Portengen, T. Ostreich and L.J. Sham, Phys. Rev. B {\bf 54}, 17452 (1996);
J.K. Freericks and V. Zlatic, Phys. Rev. B {\bf 58}, 322 (1998).
\bibitem{anderson} J.M. Robinson, Phys.Rep. {\bf 51}, 1 (1979);
P. Schlottmann,  Phys. Rev. B {\bf 22},613 (1980);
B.H. Brandow,  Phys. Rev. B {\bf 33},215 (1986);
V.Z. Vulovic and E. Abrahams,  Phys. Rev. B {\bf 36},2614(1987);
A. Houghton, N. Read and H. Wu,  Phys. Rev. B {\bf 33}, 3782 (1988);
H. Schweitzer and G. Czycholl, Sol.St.Comm. {\bf 74}, 735 (1990).
\bibitem{old} J.E. Hirsch, Phys.Rev. B {\bf 43}, 11400 (1991).
\bibitem{impurity} K. Held, C. Huscroft, R. T. Scalettar and A. K. McMahan, Phys.Rev.Lett. {\bf 85}, 373 (2000). 
\bibitem{maldague} P. Maldague, Phys. Rev. B {\bf 16}, 2437 (1977).
\bibitem{undressing} J.E. Hirsch, Phys. Rev. B {\bf 62}, 14487 (2000), 
{\bf 62 }, 14498 (2000),  and references therein.
\bibitem{marsiglio} J. E. Hirsch and F. Marsiglio, Phys. Rev. B\ {\bf 62}, 15131
(2000) and references therein.
\bibitem{fye} R.M. Fye, M.J. Martins and R.T. Scalettar,
Phys. Rev. B {\bf 42}, 6809 (1990).
\bibitem{qmc} J.E. Hirsch, Phys.Rev. B {\bf 65}, 214510 (2002).
\bibitem{hanke} J. Wagner, W. Hanke and D.J. Scalapino, Phys.Rev. B {\bf 43}, 10517 (1991).
\bibitem{last} J.E. Hirsch, cond-mat/0205006 (2002).
\bibitem{uchida} S.Uchida, T.Ido, H.Takagi, T. Arima, Y. Tokura and
S. Tajima,  Phys. Rev. B {\bf 43}, 7942 (1991).
\bibitem{marel} H. J. A. Molegraaf, C. Presura, D. van der Marel, P. H. Kes, and M. Li
Science {\bf 295}, 2239 (2002).
\bibitem{santander} A.F. Santander-Syro et al, cond-mat/0111539 (2001).
\bibitem{basov} D.N. Basov et al, Science {\bf 283}, 49 (1999).
\bibitem{tanner} F. Gao, D.B. Romero, D.B. Tanner, J. Talvacchio and 
M.G. Forrester, Phys.Rev. B {\bf 47}, 1036 (1993).
\bibitem{twoband} J. E. Hirsch and F. Marsiglio, Phys.Rev. B {\bf 43}, 424 (1991).
\bibitem{mgb2}
J. Kortus, I. I. Mazin, K. D. Belashchenko, V. P. Antropov, and L. L. Boyer,
Phys.Rev.Lett. {\bf 86}, 4656 (2001). 
\bibitem{pickett} H. Rosner, A. Kitaigorodsky and W.E. Pickett, 
Phys. Rev. Lett. {\bf 88}, 127001 (2002).
\bibitem{dynh} J.E. Hirsch, Phys. Rev. Lett. {\bf 87}, 206402 (2001).
\bibitem{hole} J.E. Hirsch, Phys.Lett. A {\bf 134}, 451 (1989).
\bibitem{parks} "Superconductivity", ed. by R.D. Parks, Marcel Dekker, New York, 1969.



\end{references}
 \end{document}